\documentclass[12pt]{article}

\usepackage{amssymb,amsmath,amsfonts,eurosym,geometry,ulem,graphicx,caption,color,setspace,sectsty,comment,footmisc,caption,pdflscape,subfig,array,hyperref, booktabs, xcolor, threeparttable, float}
\usepackage{chngcntr}

\floatstyle{plaintop}
\restylefloat{table}
\restylefloat{figure}

\usepackage[font=bf]{caption}

\usepackage[sectionbib]{natbib}

\newcolumntype{L}[1]{>{\raggedright\let\newline\\arraybackslash\hspace{0pt}}m{#1}}
\newcolumntype{C}[1]{>{\centering\let\newline\\arraybackslash\hspace{0pt}}m{#1}}
\newcolumntype{R}[1]{>{\raggedleft\let\newline\\arraybackslash\hspace{0pt}}m{#1}}

\begin{document}

\begin{titlepage}
\title{Place-Based Policies for Neighborhood Improvement: Evidence from Promise Zones}
\author{Adamson Bryant\thanks{I thank Peter Hull, Jesse Bruhn, John Friedman, Matt Turner, our AMB Research Group, and all Brown Applied Micro Lunch seminar participants for valuable feedback and suggestions. This material is based upon work supported by the National Science Foundation Graduate Research Fellowship under Grant No. 2022343505. Any opinion, findings, and conclusions or recommendations expressed in this material are those of the authors and do not necessarily reflect the views of the National Science Foundation. I am also grateful to the Population Studies and Training Center at Brown University, which receives funding from the NIH, for training support (T32 HD007338) and for general support (P2C HD041020). }}

\date{\today}
\maketitle
\begin{abstract}
\noindent Despite growing evidence that neighborhoods play a critical role in shaping economic mobility and well-being, effective policies to address neighborhood disadvantage remain elusive. This study evaluates the impact of the Promise Zone program, which aims to revitalize disadvantaged neighborhoods through streamlined federal support and grant incentives. I use an event study framework with newly obtained data on the location of failed finalist applications as a comparison group to estimate the program’s effects. The results reveal significant improvements in poverty, household incomes, and employment in Promise Zone neighborhoods, particularly in later-designated zones and initially low-status neighborhoods. I also find that effects are driven partly by changes in residential composition, and that Promise Zones appear to induce positive spillovers in adjacent areas. \\
\vspace{0in}

\bigskip
\end{abstract}
\setcounter{page}{0}
\thispagestyle{empty}
\end{titlepage}
\pagebreak \newpage

\doublespacing

\section{Introduction} \label{sec:introduction}

Neighborhoods matter: A large body of work documents the perils of growing up in poor neighborhoods \citep{wilson_truly_1987, mayer_growing_1989, sampson_assessing_2002}, and neighborhood poverty rates correlate with adult and child outcomes we care about, such as employment, life expectancy, and academic achievement \cite{chyn_neighborhoods_2021}. Importantly, there is now evidence that effects are causal for adult physical and mental health \cite{ludwig_long-term_2013}, and, most notably, for children's long term earnings \citep{chetty_effects_2016, chetty_impacts_2018, chyn_moved_2018, chyn_effects_2023, haltiwanger_children_2024}. 

Given these documented neighborhood effects, policymakers may ask what can be done to improve the lives of those residing in poor neighborhoods. There has been a long-running debate among economists about whether it is better to help people or places, with the conventional view being that we should directly subsidize these households via reforming the tax and transfer system \citep{austin_saving_2018}. Subsidizing places distorts the allocation of productive inputs, and in a classic Rosen-Roback model results in those subsidies being capitalized into land rents - thus benefiting the landowners rather than the residents \citep{rosen_housing_1979, roback_wages_1982}. However, more recent work from \cite{kline_people_2014} argue that there are a number of market failures that could provide justification for place-based policies. If the relationship between concentration of poverty and the negative neighborhood externality that it exerts on its residents is nonlinear, then helping poor people in poor places can have a multiplier effect. However, there still exists a lack of evidence-backed, effective policy tools that focus on improving neighborhoods. 

This paper studies the impact of the federal Promise Zone program, a place-based policy introduced in the second Obama administration. The program aimed to improve neighborhood living and working conditions through grant incentives and increased federal support. To study the effectiveness of the program, this paper focuses on three measures of a neighborhood's condition: the poverty rate, median household income, and employment to population ratio. Using newly obtained data from the Department of Housing and Urban Development (HUD) acquired through a Freedom of Information Act (FOIA) request, a counterfactual set of neighborhoods are constructed from the applicants honored as ``finalists" but not selected as Promise Zones. Comparing winning Promise Zone applicants (``designees") with the failed finalist applicants (``finalists") in an event study framework, I find significant improvements for neighborhoods within the designated Promise Zones. In the post-period, designee neighborhoods see 2.0 percentage points lower poverty rates, 1.6 percentage points larger employment to population ratios, and 8.1\% greater median household incomes. These effect sizes represent real but modest improvements, ranging from 4 to 8\% of baseline means. To summarize the effects across these different dimensions, I take inspiration from \cite{kling_experimental_2007} and aggregate the three primary outcomes into a Neighborhood Status Index. Designee neighborhoods see a differential increase of 0.20 standard deviation in the Neighborhood Status Index, which suggests that about 30\% of the observed improvement in these neighborhoods over the post-period was due to the Promise Zone program.  These results rely on a parallel trends assumption, which I provide confidence for by showing that the trends between the designees and finalists track closely in the pre-period. 

I conduct a number of additional analyses to further investigate the effects of this program. The Average Treatment on the Treated effects estimated in the baseline analysis mask three pieces of significant heterogeneity. First, the Promise Zone program was rolled out over three designation periods in 2014, 2015, and 2016, and the aggregate estimate is driven largely by the effects for the cohort of 2016 designees. The 2016 designees had much stronger proposals, as evidenced by the number of finalists in that round, and thus were better poised to take advantage of the Promise Zone benefits. Second, neighborhoods of initial low status by my index measure saw larger gains from the program than neighborhoods of initially higher status. This is a surprising result in the context of similar policies, such as Empowerment Zones, that had the opposite effect \citep{reynolds_effects_2015}. Third, local implementation varied considerably across sites. Using an active online presence as a proxy for local implementation quality suggests that sites with stronger implementation saw program effects that were twice as large as those with weaker implementation. 

Given the significant effects in the main analysis, a natural question to ask is whether those effects were driven by improvements of the original residents or by a change in neighborhood composition. I find conflicting evidence on this point: on one hand, the share of households making over \$100k increased dramatically, suggesting the neighborhoods became more attractive to wealthier residents; on the other hand, there are no effects on the demographic makeup of these neighborhoods, with the data rejecting larger than 2 percentage point changes in the share of non-white and single-parent households as a result of this policy. Finally, I consider the spatial aspect of this program by studying spillovers to nearby neighborhoods. Positive effects extended to nearby neighborhoods, with neighborhoods a half mile from the border seeing effect sizes that were about half of those for neighborhoods inside the border. An extrapolation exercise suggests that these positive effects extended for about a mile from the border as well. 

This paper contributes to the literature on place-based policies by providing, to the best of my knowledge, the first estimates of the effects of the Promise Zone program across all locations. It builds on two other works that study individual Promise Zones. \cite{kitchens_impact_2022} study the effects of the Los Angeles Promise Zone on the housing market using a spatial regression discontinuity design, finding that the program appreciated housing values. \cite{marsella_effect_2024} studies crime in the West Philadelphia Promise Zone, finding a significant reduction using a synthetic difference-in-difference estimator. A number of papers have studied the effects of the Empowerment Zone program, which provided inspiration for the design of the Promise Zone program, generally finding small positive effects \citep{busso_assessing_2013, hanson_local_2009,reynolds_effects_2015, neumark_enterprise_2019}. Estimates of the effectiveness of place-based policies on reducing poverty have varied considerably, to sizable in the case of HOPE VI \citep{tach_public_2017, staiger_neighborhood_2024}, modest for the New Markets Tax Credit \citep{freedman_teaching_2012}, or null for the Opporunity Zone program \citep{freedman_jue_2023}. My estimates suggest that Promise Zones reduced neighborhood poverty rates about twice as much as Empowerment Zones and the New Market Tax Credit program, but the effects were about a third of the size of the HOPE VI program over a similar time horizon. 

The rest of this paper is organized as follows. Section \ref{sec:context} provides more details on the Promise Zone program and provides comparisons to other place-based policies. Section \ref{sec:data} describes the data and the empirical strategy. Section \ref{sec:result} presents the results from the primary empirical analysis. Section \ref{sec:spillovers} describes the analysis for investigating the effects on nearby neighborhoods and present results. Section \ref{sec:mechanisms} presents evidence for possible mechanisms. Section \ref{sec:conclusion} concludes.

\section{Institutional Context} \label{sec:context}
\subsection{Overview of Promise Zone Program}

\begin{figure}[h]
    \centering
    \includegraphics[width=0.7\linewidth]{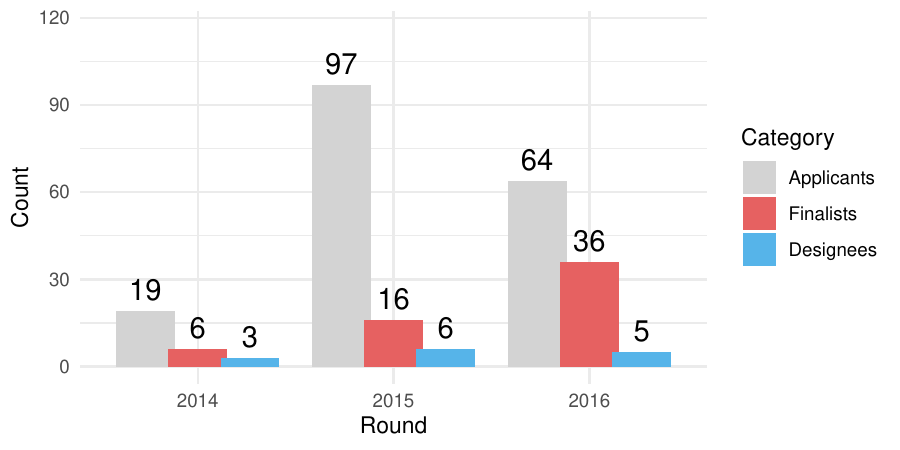}
    \caption{Overview of Promise Zone Applications by Round}
    \label{fig:apps}

    \raggedright
       {\footnotesize Notes: These categories are the total number of applications that reached each stage and are not mutually exclusive. }
\end{figure}

The Promise Zone (PZ) program was announced in 2013 during President Obama’s State of the Union address. The program was an interagency effort led by the Department of Housing and Urban Development (HUD), and there were 3 rounds of designations held in 2014, 2015, and 2016. In each round, organizations applied to one of 3 different categories of PZs: Urban, Rural, and Tribal. Due to the smaller sample sizes of Rural and Tribal categories, this paper focuses on the Urban PZs. The majority of the lead applicants were local governments, but occasionally some NGOs submitted the applications with the local government entities as co-applicants. Eligibility requirements differed slightly across the 3 rounds, but the general requirements were to be a contiguous geographic area of 10,000 to 200,000 residents, with an average poverty rate above 20\%. There were a total of 13 Urban PZs designated, and Figure \ref{fig:apps} shows an overview of the applications across each round. 



The stated scoring criteria was for ``applicants [that] demonstrated a consensus vision for their community and its residents, the capacity to carry it out, and a shared commitment to specific, measurable results" \citep{hud_promise_nodate-1}. Each application was scored out of 100 points, with applications scoring over 75 being named as ``finalists". Above 75 points, the committee ranked the applications within the categories (Urban, Rural, Tribal) and generally picked the highest ranked ones. They note that they may have picked a lower ranked application over a higher ranked application for reasons such as increasing geographic diversity \citep{hud_promise_2014}. The winners received the Promise Zone designation for an initial period of 10 years.

The idea behind the PZ program was to initiate neighborhood improvement from the bottom up, rather than top down, focusing on addressing needs outlined by local leaders. The intention was that the federal government would work closely with local leaders and provide them with necessary federal support in implementing their plan. Thus, the primary benefits were the following: 
\begin{itemize}
    \itemsep0em 
    \item An opportunity to engage five AmeriCorps VISTA members in the Promise Zone's work
    \item A federal liaison assigned to help designees navigate federal programs
    \item Preferences for certain competitive federal grant programs and technical assistance from participating federal agencies 
\end{itemize}

The AmeriCorp VISTA members are akin to domestic PeaceCorp volunteers, who work full-time for 1 year with nonprofits or government agencies and receive a living stipend. These volunteers provided extra manpower to the PZ organizations. The federal liaisons were assigned to a specific PZ and worked at HUD headquarters, providing the bridge between the local leaders and rest of the federal government. In 2014, there were 35 different grant programs across 14 agencies which the PZs received preference for \citep{hud_promise_nodate}. As a consequence of the PZ  ``bonus points", multiple local partners that had previously submitted separate grant applications often teamed up to submit a single one with the bonus points, which encouraged local collaboration \cite{zapolsky_promise_2019}. There was also proposal in Congress to enact tax benefits for PZs but it did not pass. Thus, the benefits of becoming a PZ were primarily reducing the bureaucratic costs of receiving assistance from a multitude of federal agencies. 

Because these benefits were a flexible in terms of practical application, the implementation of the PZ program appeared to vary considerably across the different zones. For example, some had monthly working group meetings, while others appeared to just send annual newsletters. This suggests there was significant heterogeneity in effects by zone.

\subsection{Comparison to Other Federal Place-Based Policies}

There are two other major ``zone" place-based policies that have been enacted by the Federal government: Empowerment Zones and Opportunity Zones. 

Promise Zones differ from the Opportunity Zone (OZ) program, which was began in 2018, in three key ways. Both had similar eligiblity requirements in terms of need, as OZs had to have poverty rates of over 20\%, but OZs were only single census tracts, rather than a set of contiguous census tracts. While PZs were selected through a competitive process, OZs were designated by state governors, with the law allowing them to designate up to 25\% of eligible census tracts in the state as OZs. The benefits of an OZ designation were very different from PZs, as an OZ designation allowed private investors to defer taxes on capital gains on business or real estate investments made within OZs, and write off the capital gains taxes completely if they held their investment for at least 10 years. There has been a significant amount of research on the short-run impacts of the OZ designations, finding mixed effects on employment and no effect on earnings or poverty rates \citep{atkins_jue_2023, freedman_jue_2023, arefeva_effect_2024}.

The Empowerment Zones (EZ) program served as the model for the Promise Zone program, with the major difference being the benefits given to designees. Like PZs, EZs were selected through a competitive process overseen by HUD, with 31 urban communities designated as EZs over three rounds from 1994-2004. EZs also had very similar eligibility criteria to PZs in terms of population size and poverty rate requirements. The EZ designation was initially supposed to last 10 years like the PZ designation, but the programs tax benefits have been continually extended by Congress up to the present day. The primary difference between the two programs is the benefits structure. An EZ designation came with employment tax credits and \$100 million in Social Services Block Grant funds. The tax credits allowed firms to receive a credit of up to 20 percent of the first \$15,000 in wages earned by each employee who lived and worked in the community. The block grants could be used for a variety of social services such as business assistance, infrastructure investment, training programs, youth services, and emergency housing assistance. The proposed tax credits for PZs did not pass, and while the PZ benefits didn't give additional block grant funding directly, the package of PZ benefits should act through a similar channel by increasing federal funding to these areas. There have been a number of studies that evaluate the effectiveness of the EZ program, generally finding an positive effects on employment and earnings, but minimal effects on poverty rates \citep{busso_assessing_2013, hanson_local_2009, neumark_enterprise_2019}. The EZ program itself was modeled after individual state Enterprise Zone policies, which had the hiring tax subsidies but not the block grants and were found to be generally ineffective \citep{neumark_enterprise_2019}. \cite{bartik_broadening_2020} speculates that this difference was due to the block grants that were included in the EZ program but not in Enterprise Zones. If this is the case, we should expect PZs to have similar effects as EZs. 



\section{Data \& Empirical Strategy} \label{sec:data}
\subsection{Data}
The unit of analysis for this study is the neighborhood, which I will follow the majority of the literature in defining as census tracts. The primary source of data comes from the 2005-2023 American Community Survey (ACS), accessed through the National Historical Geographic Information System \citep{manson_ipums_2023}, which provides demographic and socioeconomic characteristics of neighborhood residents as well as information on housing units.  The ACS estimates at the census tract level are only available as 5-year averages, so each data point is coded as the last year in that 5-year average. Tract-level data from before 2010 and after 2020 are crosswalked to the 2010 Census definition of tracts to ensure that units are geographically comparable over time. This data supplies most outcomes considered in this paper such as a neighborhood's poverty rate, median household income, and employment to population ratio. Following previous work using neighborhood-level data \citep{busso_assessing_2013}, these outcomes are combined into an index which is an equal-weighted average of each outcome normalized to a Z-score based on the distribution of treatment tracts in the baseline period. Using this index measure yields greater statistical power and allows me to sidestep concerns over multiple hypothesis testing, though it does sacrifice clear interpretability. 


The treatment status of a tract is defined by whether it falls inside a designated Promise Zone, a proposed Promise Zone from a finalist applicant, or neither. Data on the locations of designated Promise Zones were obtained from the Promise Zone Factsheets that are available from the HUD Exchange website. The locations of the proposed Promise Zones for the finalist applicants were obtained via a Freedom of Information Act (FOIA) request to the Department of Housing and Urban Development (HUD). Another FOIA request was made to obtain the scores given to each designee and finalist application, although the scores for the designee applications are pending a further follow up request. Further details on the data construction can be found in Appendix \ref{app:data}. 

Summary statistics are in Table \ref{tab:summary_stats}. The designee tracts are very poor, with average poverty rates of 34\%, and high rates of joblessness with only 40\% of residents employed. The finalist tracts are similar, although they are worse off than the designee tracts along the main outcomes of interest. Nearby tracts are defined here as tracts within 1 mile of a PZ border but 75\% or more outside the zone by area. Without access to the finalist zones as a comparison group, a common strategy to form counterfactuals for the designee tracts would be to use nearby tracts. This table shows that the finalist tracts are considerably closer to the designee tracts in terms of baseline covariates than the nearby tracts. It also shows that the Promise Zones were worse off than their immediate neighboring tracts. 

\begin{table}[h]
\begin{threeparttable}
    \centering
    \begin{tabular}{lccc}
        \hline
        & \textbf{Designee Tracts} & \textbf{Finalist Tracts} & \textbf{Nearby Tracts} \\
        \hline
        Neighborhood Status Index & 0.16 (0.83) & -0.20 (0.94) & 1.01 (1.04) \\

        Med HH Income (\$1000) & 28.8 (7.65) & 27.0 (9.73) & 41.0 (17.3) \\
        Poverty Rate (\%) & 0.34 (0.12) & 0.37 (0.13) & 0.24 (0.14) \\
        Emp-Pop Ratio (\%) & 0.40 (0.09) & 0.35 (0.08) & 0.46 (0.11) \\
        HS Dropout Rate (\%) & 0.11 (0.13) & 0.13 (0.14) & 0.09 (0.13) \\
        \% Single Parent HH & 0.35 (0.13) & 0.36 (0.11) & 0.25 (0.13) \\
        \% Nonwhite & 0.85 (0.20) & 0.85 (0.20) & 0.67 (0.31) \\
        \% College Degree & 0.13 (0.11) & 0.10 (0.08) & 0.25 (0.21) \\
        Unique Tracts & 326 & 409 & 514 \\
        \hline
    \end{tabular}
    \caption{Summary Statistics by Tract Type}
    \label{tab:summary_stats}
    \begin{tablenotes}
    \small
    \item Notes: Summary statistics calculated from 2006-2010 ACS estimates as weighted averages of tracts by population. Nearby tracts are those within 1 mile of a PZ border but 75\% or more outside the zone by area. Standard deviations are in parentheses. 
    \end{tablenotes}
\end{threeparttable}

\end{table}

\subsection{Empirical Strategy}

The baseline empirical approach is motivated by the ideal experiment in which distressed urban neighborhoods are randomly assigned the Promise Zone designation or not. However, the assignment of this program to areas was decidedly nonrandom - typically the top-scoring applications received the designation. The approach here is thus to find a suitable group of control neighborhoods to serve as counterfactuals for the Promise Zone designees. Given the competitive application assignment mechanism, the highest scoring non-winners provide a natural comparison group for the designees. I employ an event study framework which utilizes the neighborhoods in the finalist zones as a control group for the neighborhoods in the designated Promise Zones. Since there were three different designation periods, this context falls under the staggered adoption setting.

My target parameter is a weighted average ATT following the \cite{wing_stacked_2024} stacked event study approach:  $$\beta=\frac{1}{4}\sum_{e=4}^7 \sum_a \omega(a)ATT(a,e)$$ where $a\in\{2014, 2015, 2016\}$ denotes the adoption year, $e$ is event time with $e=0$ set to the time of adoption, $\omega (a)$ are population weights\footnote{The weights are defined as $\omega(a)=\frac{Pop_a^D}{Pop^D}$ where $Pop_a^D$ is the 2010 population in treated areas in adoption group $a$, and $Pop_D$ is the 2010 population of all treated areas. While all census tracts are designed to have similar population, the boundaries of the zones don't always align with census tract boundaries. The weights of those tracts not wholly contained within zones are thus based off of the population of those tracts within the zone boundaries.}, and $ATT(a,e)$ is the Average Treatment Effect on the Treated for adoption group $a$ in event time $e$. This estimand aligns with those from alternative approaches to the staggered adoption setting such as \cite{sun_estimating_2021}, \cite{callaway_difference--differences_2021}, and \cite{borusyak_revisiting_2024}; results from these approaches are shown in the Appendix, and are very similar to my baseline estimates. The \cite{wing_stacked_2024} specification builds on the stacking approach popularized in \cite{cengiz_effect_2019} by including specific regression weights based on the share of treated units from each different adoption time group in each event time window in order to correctly target the ATT estimand in stacking designs. 

The aggregation in this ATT is only over event time periods 4 through 7 due to the ACS data only coming in 5-year averages at the census tract level. Because each data point is coded as the final year in that 5-year average, it's not until $e=4$ that the average only contains underlying data from the post period. These ``partial treatment estimates" from $e=0$ through $e=3$ are grayed out in the event study figures shown in the results. It's possible to leverage additional assumptions to investigate the treatment dynamics at the year-level, which is done in Appendix \ref{sec:dynamics}. 

Given the target estimand $\beta$ above, my primary static empirical specification is \begin{equation}\label{eq:static}
    Y_{iae}=\beta (D_i \cdot 1[e\in[4,7]]) + \sum_{t=-5,\neq-1}^3 \gamma_k(D_i\cdot1[e=t]) + D_i+\delta_e +  \varepsilon_{iae}
\end{equation}
with regression weights $$Q_{iae}=\begin{cases}
      \frac{Pop_a^D}{Pop^D}/\frac{N_a^D}{N^D}, & \text{if}\ D_i=1 \\
      \frac{Pop_a^D}{Pop^D}, & \text{if}\ D_i=0
    \end{cases}$$
where $Y_{iae}$ denotes the outcome for tract $i$ in adoption group $a$ at event time $e$, $D_i=1$ if the tract is in a designated Promise Zone, $1[e\in[4,7]]$ is an indicator for whether the event time is fully in the post-period, $1[e=t]$ is an indicator for whether the event time is equal to $t$, and $\delta_e$ is an event time fixed effect. The weights adjust for the population in tracts across each adoption group, with $Pop_a^D$ is the 2010 population in treated areas in adoption group $a$,  $Pop_D$ is the 2010 population of all treated areas, $N_a^D$ is the number of designee tracts in adoption window $a$, and $N^D$ is the total number of designee tracts. The specification above ensures that the estimated coefficient $\hat{\beta}$ above corresponds to my target estimand. Due to the spatial nature of the data, inference is conducted using \cite{conley_gmm_1999} standard errors that allow for decaying spatial autocorrelation between units, in addition to allowing for autocorrelation over time within units. The baseline approach uses a 10 mile cutoff for the spatial kernel, and results are not sensitive to this choice. To get a sense of dynamics and investigate parallel trends in the pre-period, I also run the dynamic specification
\begin{equation}\label{eq:dynamic}
    Y_{iae}=\sum_{t=-5,\neq-1}^7 \gamma_k(D_i\cdot1[e=t]) + D_i+\delta_e +  \varepsilon_{iae}
\end{equation} with the same $Q_{iae}$ weights as above.

With this approach to identification, the two standard event study assumptions, no anticipation and parallel trends, need to be made in order to identify the ATT. It's not possible to directly test these two assumptions. The typical approach to build confidence in these assumptions is to plot event study coefficients $\gamma_k$ for the periods before treatment, which is done in Figure \ref{fig:main_nsi}. A benefit of the stacking approach to staggered adoption is that this visual analysis can be interpreted in the same way as the familiar dynamic two-way fixed effects approach in the non-staggered adoption case, avoiding issues raised by \cite{roth_interpreting_2024}.  Since this program was only announced a year before the first round of applications began and the application process was extremely competitive (over the 3 rounds, only 14 designees were selected out of 120 applications), it seems very unlikely that there would have been effects of this program before it was enacted. Under the no anticipation assumption, pre-period coefficients of 0 imply that the parallel trends assumption holds in the pre-period. This builds confidence that the parallel trends would have continued into the post-period if not for the treatment, which then attributes the differential effects observed in the post-period to treatment effects. As noted by \cite{roth_pretest_2022}, a non-rejection test of pre-period parallel trends has quite low power. I thus follow \cite{rambachan_more_2023} by conducting sensitivity tests to pre-trend violations. In particular, I report confidence intervals that account for the possibility of linear pre-trends that fall within the confidence intervals of the estimated pre-period coefficients. 

\section{Results} \label{sec:result}
\subsection{Main Results} \label{ssec:res_main}

\begin{figure}[h]
    \centering
    \includegraphics[width=0.7\linewidth]{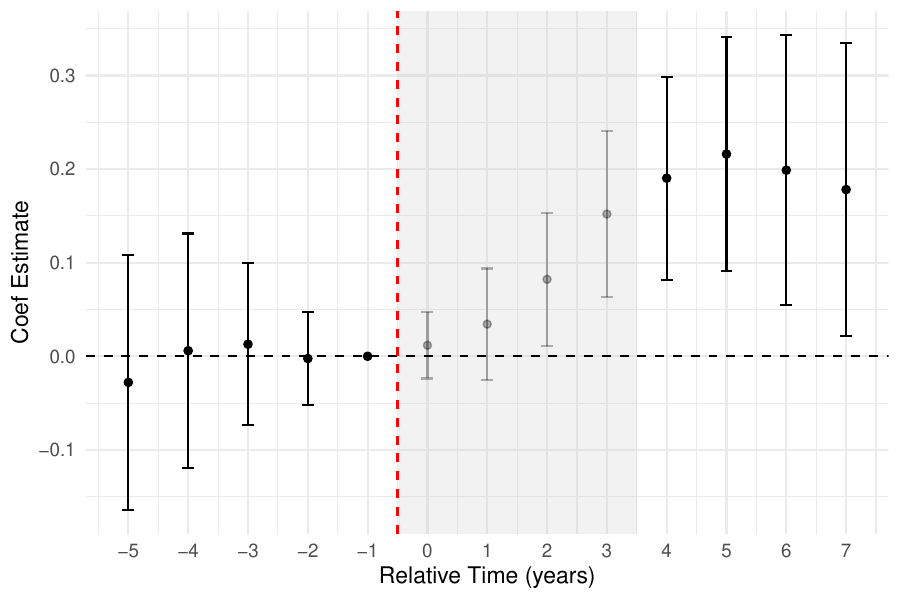}
    \caption{Baseline Specification on Neighborhood Status Index}
    \label{fig:main_nsi}
    \raggedright
       {\footnotesize Notes: Event study coefficients from baseline empirical specification on Neighborhood Status Index using \cite{conley_gmm_1999} standard errors. Grayed out areas correspond to partial treatment estimates.}
\end{figure}

The main result is plotted in Figure \ref{fig:main_nsi}. Before the Promise Zone designations, the finalists and eventual designees had very similar trends, but after the designations were made, the designees improved significantly. The grayed out area is due to those estimates being comprised of ``partial treatment" years since each point corresponds to an average over the previous 5 years. A version where those first 4 points have been inflated to account for this partial treatment is shown in Appendix Figure \ref{fig:main_nsi_scaled}. It suggests that the program had increasing effects over the first few years that stabilized starting in year 3. The ATT estimate constructed over the 4 coefficients drawing from only post-period years is shown in Table \ref{tab:main_res}, which indicates that the Promise Zone designation improved neighborhoods in the designated Promise Zones by about one-fifth of a standard deviation. Over the 7 years since the program began, the designee neighborhoods saw an 0.66 average increase in their NSI measure. This suggests  about 30\% of that increase was due to a causal effect of the Promise Zone program. 

\begin{figure}[h]
    \centering
    \subfloat[Poverty Rate \label{fig:main_pov}]{\includegraphics[width=0.33\linewidth]{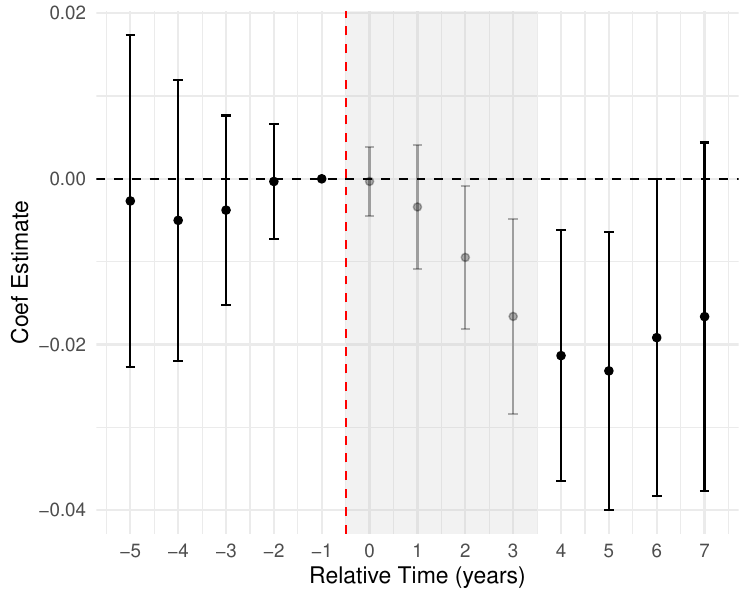}}
    \subfloat[Med HH Income (log) \label{fig:main_inc}]{\includegraphics[width=0.33\linewidth]{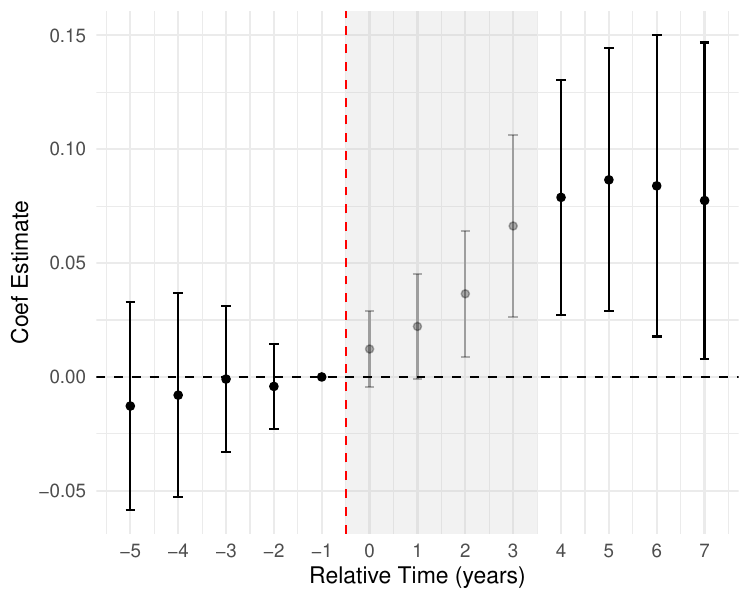}}
    \subfloat[Emp Pop Ratio \label{fig:main_emp}]{\includegraphics[width=0.33\linewidth]{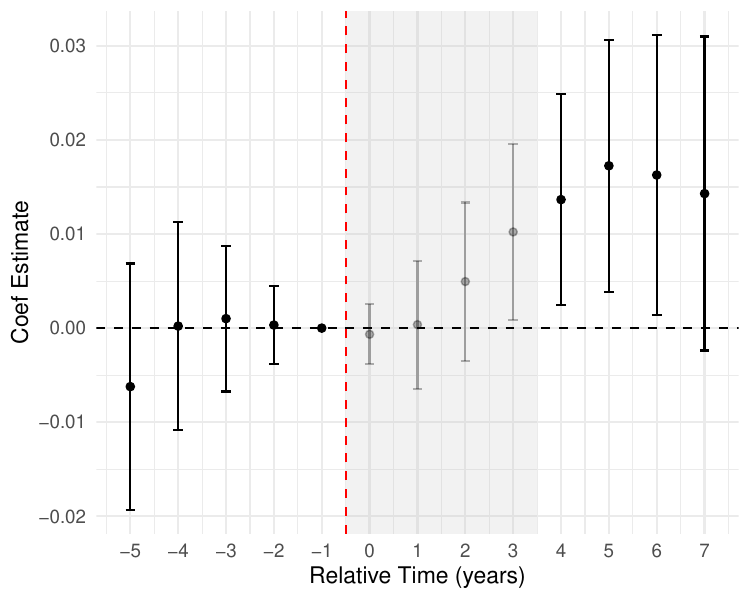}}

    \caption{Baseline Specification Event Study Plots}
    \label{fig:main_plots}

    \raggedright
       {\footnotesize Notes: Event study coefficients from baseline empirical specification on different outcomes using \cite{conley_gmm_1999} standard errors. }
\end{figure}

To decompose the index into more interpretable changes, the event study plots for the three primary outcomes that encompass the index are in Figure \ref{fig:main_plots}. Panel (a) shows that the designee neighborhoods had slightly increasing poverty rates before the Promise Zone designations took place, but the trend then reversed after the time of designations. Panels (b) \& (c) show difference in trends that are minimal before designation, but then diverge as  the designee neighborhoods show improvement. The ATT estimate for each of the individual outcomes are in Table \ref{tab:main_res}. The ATT results indicate that the Designee neighborhoods experienced a 2.0(se = 0.9) percentage point decrease in poverty, 8.1(se = 3.1) percent increase in median household income, and 1.6(se =0.7) percentage point increase in their employment to population ratio relative to finalist neighborhoods. To put these magnitudes into context, \cite{busso_assessing_2013} find that the Empowerment Zone program increased average weekly wages of residents by 5 percentage points, and \cite{neumark_enterprise_2019} estimate that the Empowerment Zone program reduced poverty rates by 1.4 percentage points. 

\begin{table}[h]
\begin{threeparttable}
    
    \centering
    \begin{tabular}{lcccc}
        \hline
        & NSI & Poverty Rate & Med HH Income & Emp-Pop Ratio \\
        \hline
        ATT Estimate & 0.198(0.067) & -0.020(0.009) & 0.081(0.031)& 0.016(0.007)\\
        p-value & 0.00 & 0.03 & 0.01 & 0.02\\
        Baseline Mean & 0 & 0.37 & \$27k & 0.39\\
        Pre-trends robust p & 0.23 & 0.29 & 0.26 & 0.36 \\
        Pre-trends robust 95\% CI & (-0.09,0.36) & (-0.06, 0.02) & (-0.02, 0.11) & (-0.013, 0.037)\\
        \hline
    \end{tabular}
    \caption{ATT Estimates in Primary Event Studies}
    \label{tab:main_res}
    \begin{tablenotes}
    \small
    \item Notes: ATT estimates calculated over event times 4-7. \cite{conley_gmm_1999} spatial kernel standard errors with 10 mile cutoff in parentheses. Pre-trends robust p-value and confidence intervals come from linear smoothness restrictions approach in \cite{rambachan_more_2023}. 
    \end{tablenotes}
    \end{threeparttable}

\end{table}

Table \ref{tab:main_res} shows 95\% confidence intervals as well as p-values from the \cite{rambachan_more_2023} formalization of the ``straight line test" for a parallel trends violation. The p-values from these tests indicate the $\alpha$ value such that a pre-trend through all $\alpha$ confidence intervals could not ``explain away" the results. The realization of these values indicate that the main results are sensitive to a pre-trend violation, a consequence of the wide pre-period confidence intervals. There have been a number of different solutions proposed in recent years to the staggered adoption difference-in-differences setting. Appendix Table \ref{tab:sun_abe} shows the coefficients estimated using a number of different estimators, which all return qualitatively the same results as my primary specification. One may be worried that differential local shocks could be driving these results. Appendix \ref{sec:local_comps} uses a Synthetic Difference-in-Difference estimator \citep{arkhangelsky_synthetic_2021} to construct comparison groups for the designees from local tracts between 2 and 10 miles away from the Promise Zone border. While this estimator is imprecise, the point estimates suggest similar conclusions to the ones made from the primary specification that use the finalists as control groups. 

\subsection{Heterogeneity}

To gain a better understanding of the program's effects, this section covers results from three informative dimension of heterogeneity: by adoption period, by initial neighborhood status, and by local implementation. For simplicity, they focus on the Neighborhood Status Index outcome. This section also addresses concerns over concurrent place-based policies.

\subsubsection{Adoption Period}

\begin{figure}[h]
    \centering
    \subfloat[2014 Designees \label{fig:het_index_2014}]{\includegraphics[width=0.33\linewidth]{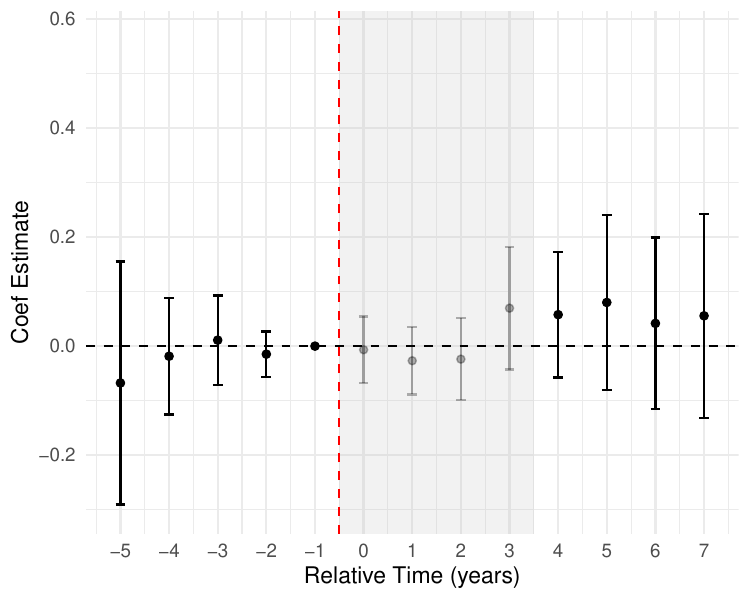}}
    \subfloat[2015 Designees \label{fig:het_index_2015}]{\includegraphics[width=0.33\linewidth]{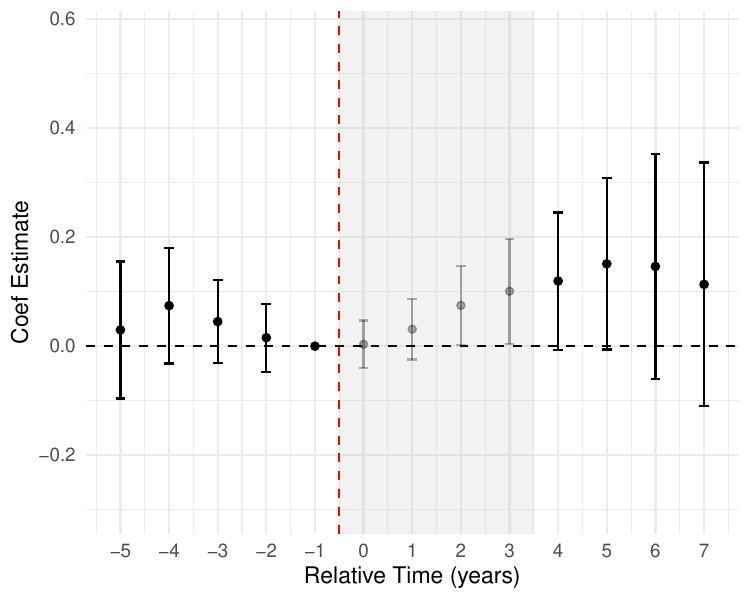}}
    \subfloat[2016 Designees \label{fig:het_index_2016}]{\includegraphics[width=0.33\linewidth]{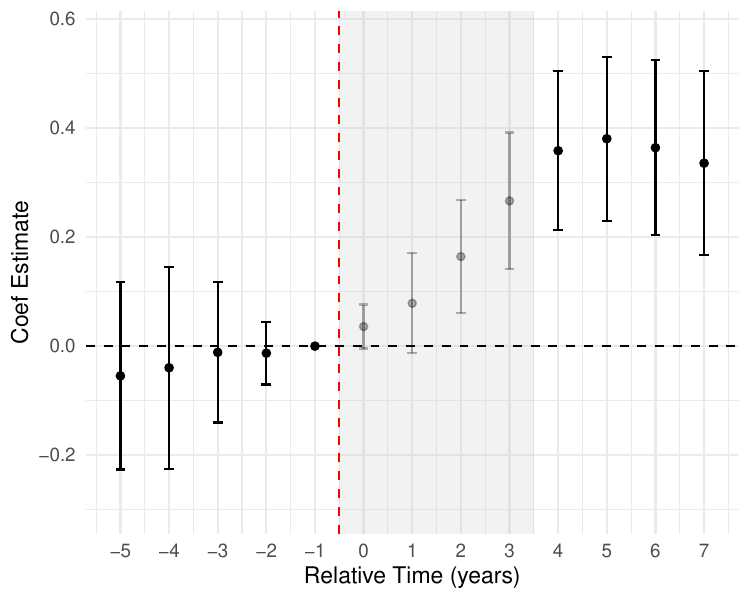}}

    \caption{NSI Treatment Effects in different Designee Cohorts}
    \label{fig:het_adoptionyear}

    \raggedright
       {\footnotesize Notes: Event study coefficients from baseline empirical specification on Neighborhood Status Index using \cite{conley_gmm_1999} standard errors. Each panel is run with all controls and a restricted treatment group}
\end{figure}

There were 3 different rounds of designation, 2014, 2015, and 2016 which had slightly different application and eligibility criteria. Event study plots for the analyses that use the full control group but the 3 different treatment groups are shown in Figure \ref{fig:het_adoptionyear}. 

These plots show that the treatment effect was greater in the later years, with almost the entire main effect being driven the 2016 designees. The likely reason for this is that the designees in 2016 had higher quality plans for neighborhood improvement than the ones in earlier years. Applications were scored primarily on the plans to improve the neighborhood and the capacity of local implementation partners, and, as mentioned earlier, applications above the threshold of 75 points were considered ``finalists". Therefore, the applicants were essentially being scored on their predicted treatment effect. As seen in Figure \ref{fig:apps}, there were 6 finalists for 3 spots in 2014, 16 for 6 spots in 2015, and 36 for 5 spots in 2016, showing that the cohort of 2016 designees were drawn from a pool of stronger applicants and were likely to be stronger candidates.

\subsubsection{Tracts by initial Status}

\begin{figure}[h]
    \centering
    \subfloat[Low Initial Status \label{fig:het_index_low}]{\includegraphics[width=0.5\linewidth]{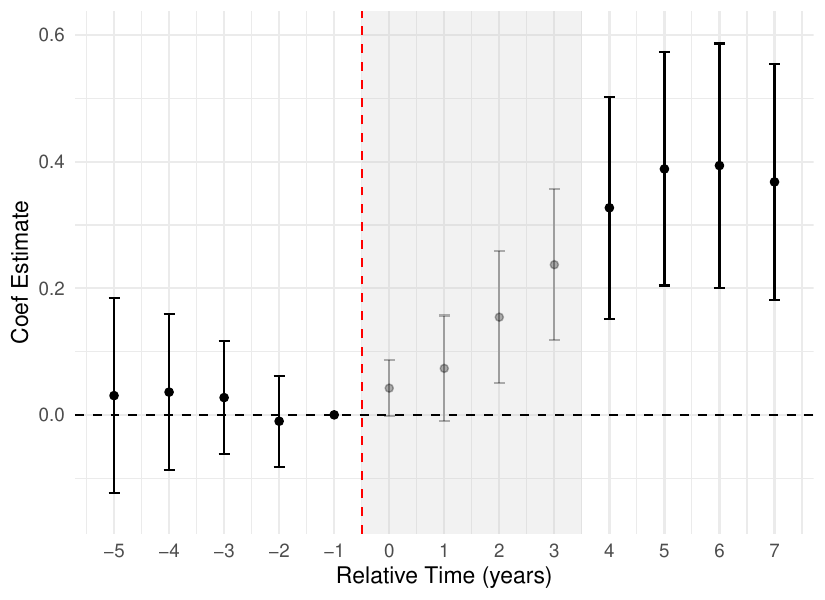}}
    \subfloat[High Initial Status \label{fig:het_index_high}]{\includegraphics[width=0.5\linewidth]{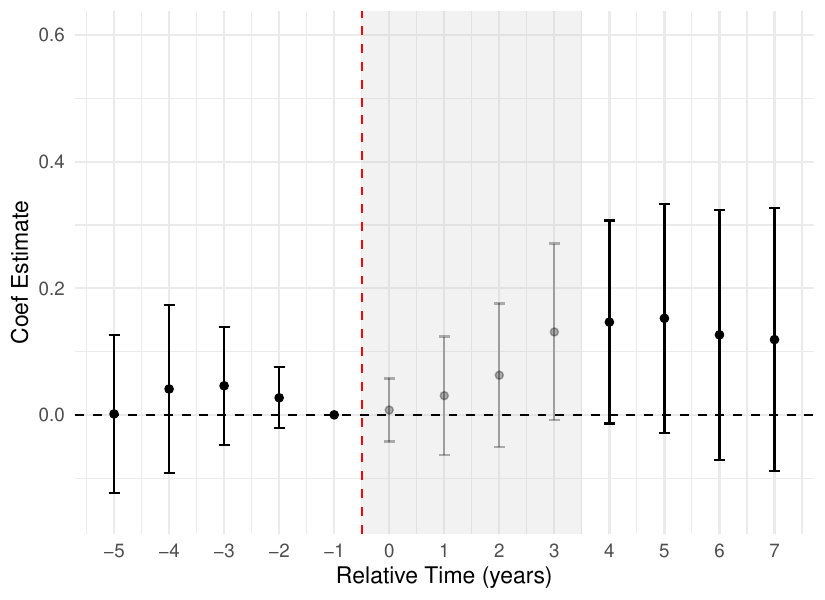}}
    
    \caption{NSI Treatment Effects by Baseline Status}
    \label{fig:het_initialstatus}
    \raggedright
       {\footnotesize Notes: Event study coefficients from baseline empirical specification on Neighborhood Status Index using \cite{conley_gmm_1999} standard errors. Low Initial Status defined by NSI values below 0 in the baseline period and High Initial Status above 0. Both treatment and control group restricted to be Low/High Initial Status.}
\end{figure}

Figure \ref{fig:het_initialstatus} shows the event study plots when restricting both the treatment and control groups to only initially low status neighborhoods, those with NSI values below 0 in the baseline year, or to initially high status neighborhoods, those with NSI values above 0 in the baseline year.

These results demonstrate that there was a much larger effect of this policy on initially low-status neighborhoods than initially high-status neighborhoods. This result points to a success of the program whose goal was to revitalize the most disadvantaged neighborhoods. This result stands in contrast to the findings from \cite{reynolds_effects_2015}, who find that in Empowerment Zones the treatment effects were driven by the neighborhoods that were initially better off. 

Additionally, Appendix Figure \ref{fig:het_surround} shows that this result holds when surrounding neighborhoods are above and below median as well. This is partially consistent with \cite{diamond_who_2019}, who find that buildings financed by the Low Income Housing Tax Credit have positive effects in low-income areas but negative effects in high-income areas. In this case, the effects are much larger in low-income areas, but they still appear to be non-negative in higher income areas. 

\subsubsection{Local Implementation}

The on-the-ground implementation of the PZ program varied widely across the different sites: some had monthly meetings across multiple working groups attended by dozens of partners, and some appeared to just submit annual reports to the city. While it's difficult to quantify the amount of resources the city government and local partners put into their PZ organizations, one measurable variable is whether the Promise Zone had an active online presence throughout its designation. These updates often were advertising events or available resources, showing that the PZ organization was actively working to help the community. Since the 2016 designees are still on their 10th and final year of PZ designation, ``active online presence" is measured by whether the PZ was actively updating their website or social media in the 9th year after designation. This excludes places that only posted their annual newsletter. By this definition, 7 of the 14 PZs are categorized as having an active online presence. 

\begin{figure}[h]
    \centering
    \subfloat[Not Active Online \label{fig:het_online_0}]{\includegraphics[width=0.5\linewidth]{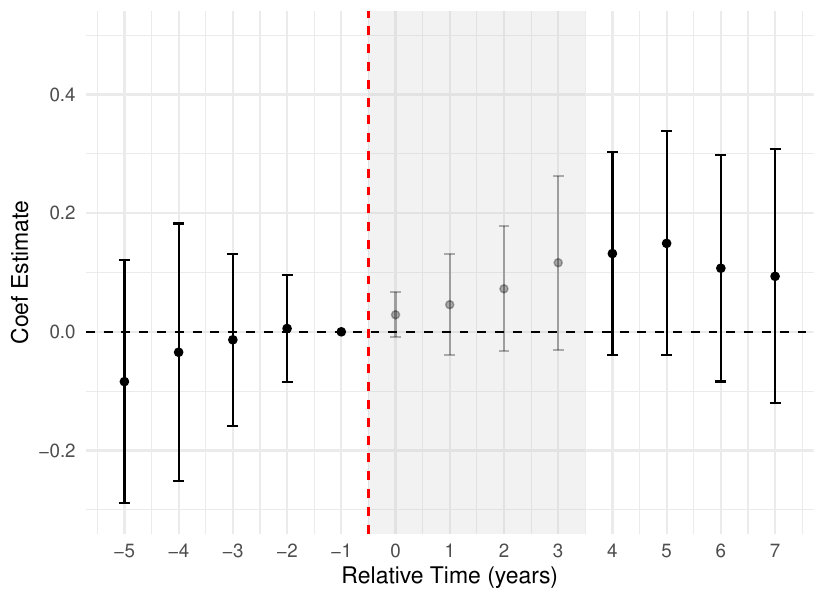}}
    \subfloat[Active Online Presence \label{fig:het_online_1}]{\includegraphics[width=0.5\linewidth]{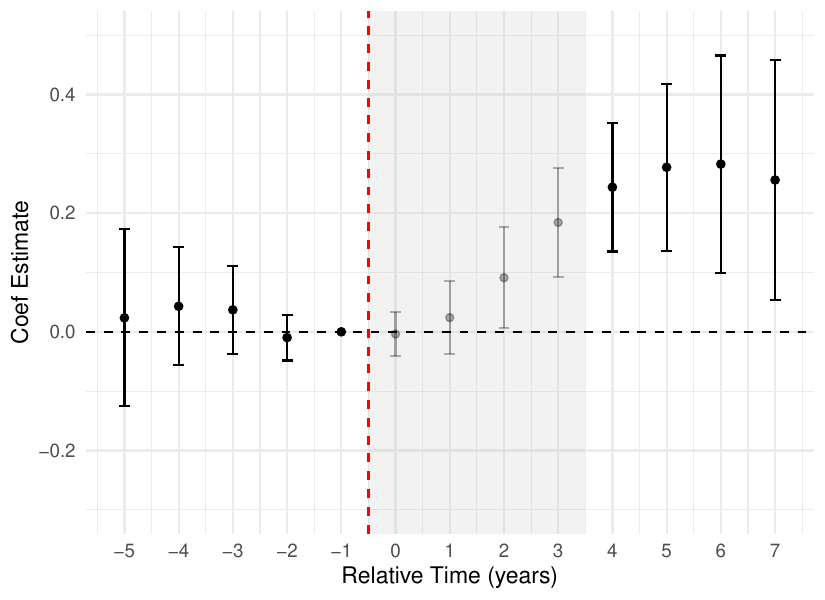}}
    
    \caption{NSI Treatment Effects by Active Online Presence}
    \label{fig:het_online}
    \raggedright
       {\footnotesize Notes: Event study coefficients from baseline empirical specification on Neighborhood Status Index using \cite{conley_gmm_1999} standard errors with different treatment groups. Promise Zones considered to have active online presence if they were updating their website or posting on social media in the 9th or later year of their designation, excluding purely posting annual newsletters. By this definition, the online PZs were Los Angeles, West Philadephia, Indianapolis, North Hartford, Sacramento, South Los Angeles, and San Diego. The non-online PZs were San Antonio, Camden, Minneapolis, St Louis, Atlanta, Nashville, and Evansville.}
\end{figure}

Figure \ref{fig:het_online} shows the results from running the event-study specification for using only designees that were either active online or not active online. The treatment effects were much larger for the neighborhoods in these PZs that were active online, even ignoring the potential parallel trends violations that would amplify this difference. Interacting a dummy for active online presence with the main specification, while underpowered, suggests that the active PZs had an ATT of 0.14(se = 0.11) standard deviations of the neighborhood status index greater effect than the non-active PZs. This is more than double the estimated treatment effect of 0.12(se = 0.10) for just the non-active PZs. These results suggest that this program's success depends on neighborhood support and effective local implementation.

\subsubsection{Concurrent place-based policies}

One concern when evaluating place-based policies is that many of them target the same areas, distressed urban neighborhoods, and thus one may be incorrectly attributing changes to one policy versus another. There is minimal overlap between Promise Zones and Empowerment Zones, with only 5\% of control neighborhoods and 10\% of treated neighborhoods belonging to Empowerment Zones. Furthermore, the Empowerment Zone program began in 1994 and the last designations were made in 2002, more than 10 years prior to the first Promise Zone designations. While the Empowerment Zone tax credits continue to be retroactively extended, these effects didn't differentially affect the treatment and control groups in the pre-period, so it seems unlikely they would have had an effect in the post-period\footnote{Similarly, Renewal Communities had an overlap of 19\% in the finalist group and 10\% in the treatment group, but their benefits began in 2002. Renewal Communities received employment tax credits like Empowerment Zones but no block grants, and the tax credits were half the amount.}.
On the other hand, there is significant overlap between Promise Zones and Opportunity Zones, which were designated in 2018, just 2-4 years into the post-period for these groups. A total of 33\% of finalist neighborhoods became Opportunity Zones, compared to 43\% of designee neighborhoods. Appendix Figure \ref{fig:het_oz} shows the event study plots from running the main specification separately for the tracts that became Opportunity Zones and the ones that did not. These figures show the same treatment effect for the two groups, and fully interacting an Opportunity Zone dummy variable with the main specification yields a treatment effect heterogeneity estimate for the ATT coefficient of 0.03(se = 0.20). Additionally, a specification that controls for event time interacted with an Opportunity Zone dummy yields only dilutes the main Promise Zone ATT effect to 0.18(se = 0.06). Thus, the introduction of Opportunity Zones are not driving the estimated Promise Zone effects.

\section{Spillovers} \label{sec:spillovers}

Classic models of spatial equilibrium raise the concern of place-based policies being zero-sum: if the targeted area benefits, then those benefits necessarily come at the expense of another area. People and businesses can move to take advantage of the benefits afforded by the policy, thereby harming the places they came from. The ``losers" here are likely to be the nearby neighborhoods, since moving costs increase with distance. In the context of Empowerment Zones, there is evidence that businesses relocated from nearby areas outside the zone to inside the zone in order to take advantage of the benefits \citep{hanson_spatially_2013}. 

However, place-based policies could also have positive effects on the adjacent neighborhoods. If low-status neighborhoods exert negative externalities on their residents, they likely also exert negative externalities on their proximate neighborhoods as well. If there is improvement in these low-status neighborhoods, then this should weaken the negative externality exerted on adjacent neighborhoods. 

To empirically evaluate this question, this section exploits the fact that we can use the finalist zones to create counterfactuals not only for tracts within Promise Zones, but also in the areas surrounding them. Figure \ref{fig:spill_map} gives a visual depiction of the comparisons that can be made. It shows a map of census tracts in Los Angeles, which contains both the Los Angeles Promise Zone (designated in 2014) and the South Los Angeles Promise Zone (designated in 2016). Tracts that overlap 75\% or more with the Promise Zones are classified as inside, and ones with between 25\% and 75\% overlap are classified as border tracts. The remaining tracts have less than 25\% overlap, and are classified based on the distance of the tract centroid from the Promise Zone border. The main analysis in this paper compared tracts between designee zones and finalist zones on the inside of the Promise Zone borders, but we can also run event studies for the bordering tracts and the tracts just outside the border of the designee and finalist zones.

\begin{figure}
    \centering
    \subfloat[Inside Tracts \label{fig:spill_inside}]{\includegraphics[width=0.33\linewidth]{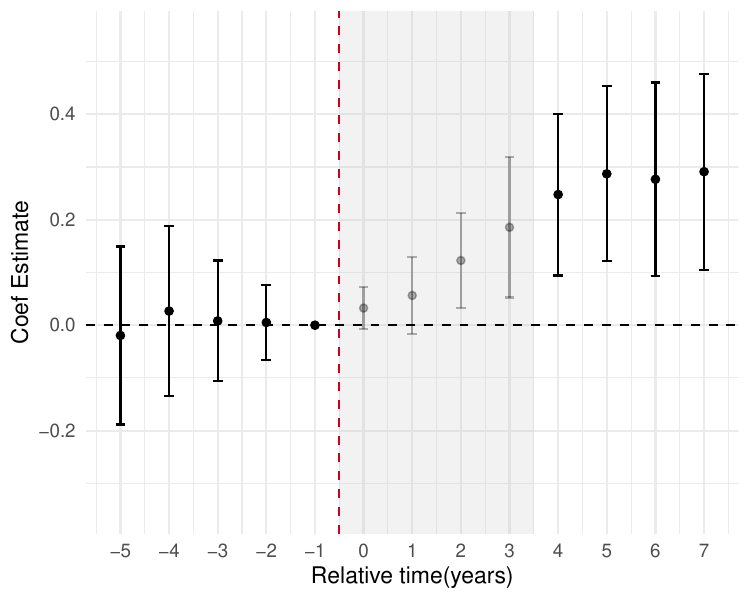}}
    \subfloat[Border Tracts \label{fig:spill_border}]{\includegraphics[width=0.33\linewidth]{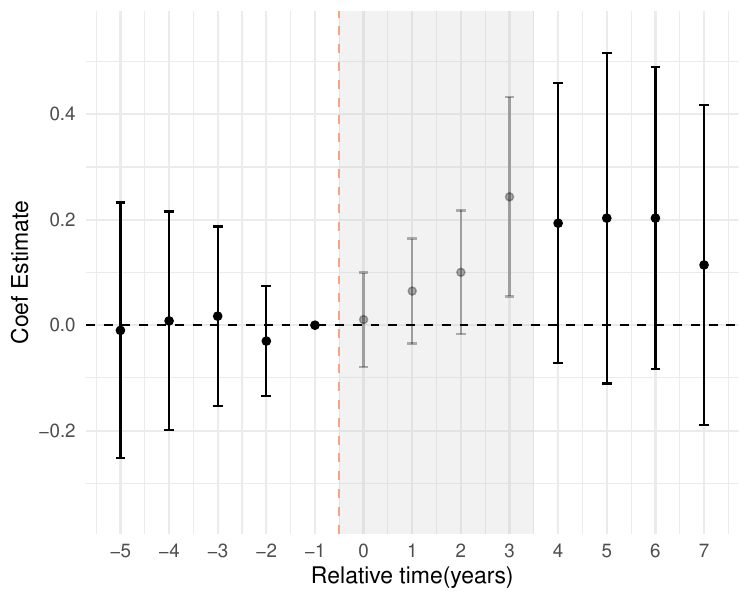}}
    \subfloat[Tracts within 1 mile \label{fig:spill_outside1}]{\includegraphics[width=0.33\linewidth]{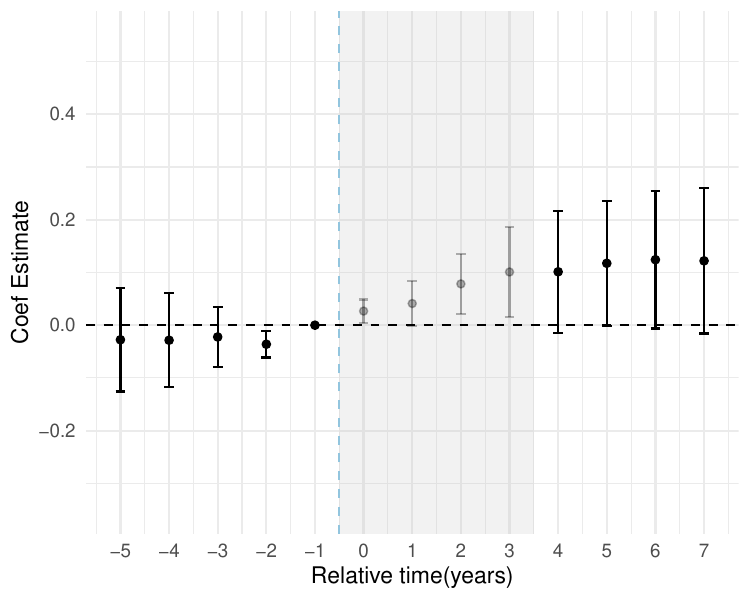}}
    \caption{Event Study plots for Different Spillover Groups}
    \label{fig:spill_eventstudy}
    \raggedright
       {\footnotesize Notes: Event study coefficients from baseline empirical specification on Neighborhood Status Index using \cite{conley_gmm_1999} standard errors. Inside tracts defined as being 75\% or more inside the zone borders. Border tracts defined as being between 25\% and 75\% inside the zone borders. Outside tracts defined as being 75\% or more outside and centroids within 1 mile of zone borders.}
    
\end{figure}

Results of this analysis are shown in Figure \ref{fig:spill_eventstudy}. Panel (a), corresponding to inside tracts,  replicates the main results from section \ref{sec:result}\footnote{The main results use all tracts with some overlap, weighting by the share of population in the zone using 2010 Census-block data. For more details see the Appendix}. Panels (b) and (c) show that positive effects extended past the interior tracts into neighborhoods at and near the border of the Promise Zones, with the impact diminishing for groups further away from the zone boundaries. Appendix Figure \ref{fig:spill_decay} takes the point estimates and graphs the effect sizes against the average neighborhood distance from the border, visualizing the spatial decay. A linear extrapolation from the point estimates suggests that the program had positive effects up to about a mile from the Promise Zone border. These estimates are consistent with the Promise Zone program reducing the negative externality of the interior neighborhoods on the geographically proximate neighborhoods. 

\section{Mechanisms} \label{sec:mechanisms}

The results from the primary analysis indicate that the Promise Zone program improved the neighborhoods it targeted. There are two channels through which these improvements could have occurred. The first is that the program improved the status of the residents initially in these neighborhoods, leading to improvement in the overall characteristics measured at the neighborhood level. The other possibility is this program made the neighborhoods more attractive to live in for employed and higher-income individuals, meaning the improvements were driven by a change in composition.

\subsection{Composition}
\label{ssec:comp}

This section studies the change in composition channel by implementing two approaches: studying the full distributions of incomes, and studying fixed demographic characteristics that should be unchanged by this policy. 

\subsubsection{Income Distribution}

\begin{figure}[h!]
    \centering
    \subfloat[Designee Distribution]{\includegraphics[width=0.45\textwidth]{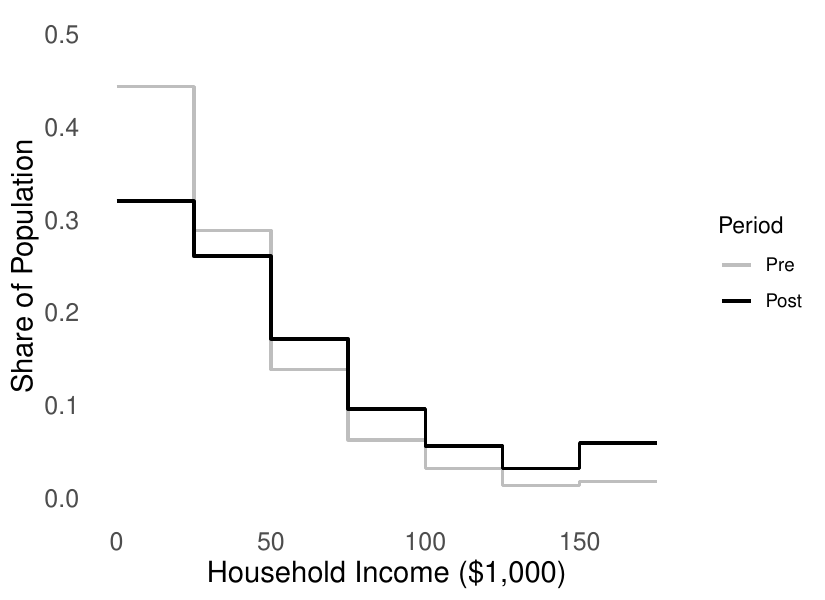}}
    \hfill
    \subfloat[Finalist Distribution]{\includegraphics[width=0.45\textwidth]{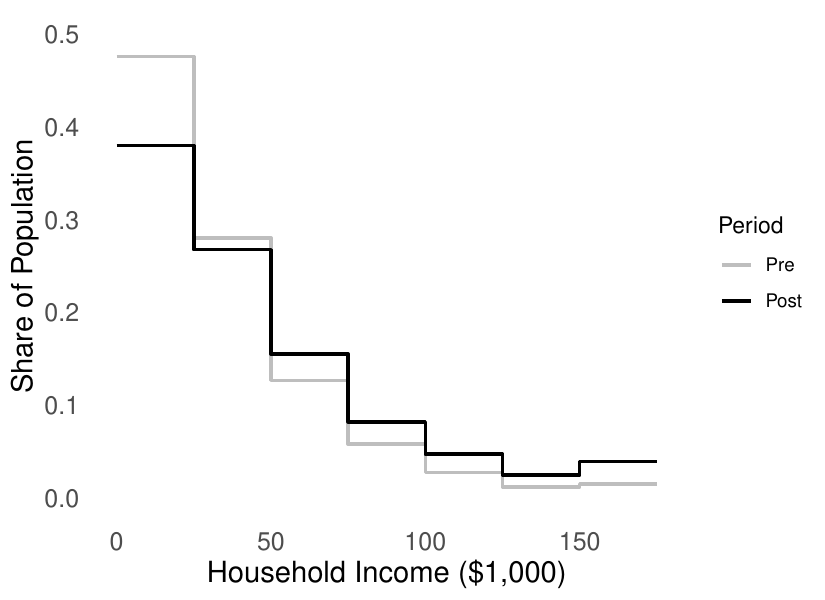}}
    
    \vspace{0.5cm}
    
    \subfloat[Treatment Effects across Distribution]{\includegraphics[width=0.6\textwidth]{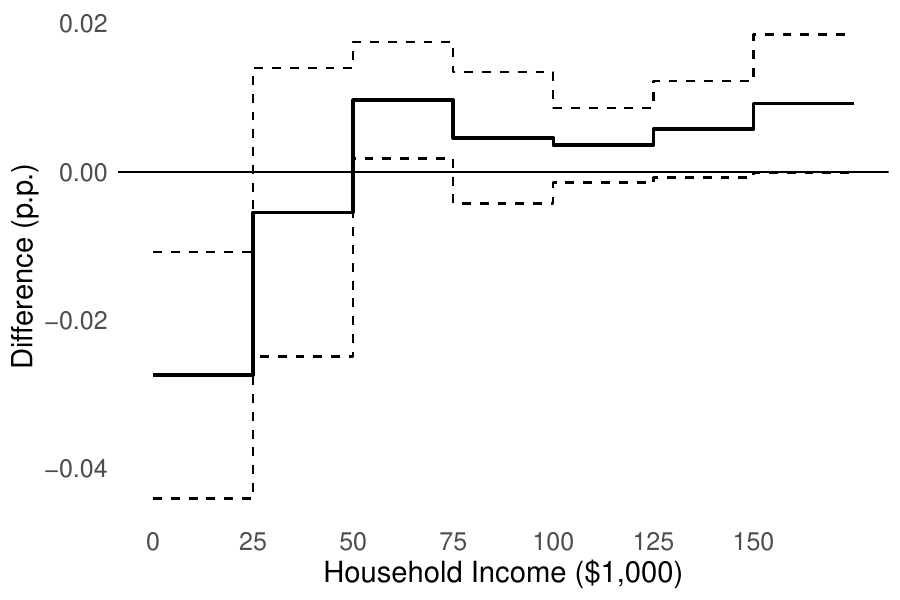}}
    
    \caption{Effects on Relative Shares of Income Distribution}
    \label{fig:comp_inc_dist}

    \raggedright
       {\footnotesize Notes: Panels (a) and (b) show income distribution from pre-period($e=-1$) and post-period($4\leq e \leq$ 7) that are formed as population-weighted averages of the respective groups. Panel (c) plots estimated ATTs and 95\% confidence intervals from the baseline specification on shares of the income groups plus a correction for inflation owing to the differing baseline distributions. All incomes above \$150k are in the final bucket.}
\end{figure}

Panels (a) and (b) of Figure \ref{fig:comp_inc_dist} plot the income distributions of the Designee and the Finalist neighborhoods in the baseline period and the post-period. Both groups of neighborhoods saw improvement over time with smaller shares of residents in the two lowest income groups, although this difference appears larger in the Designee neighborhoods. Panel (c) formalizes that observation by plotting the causal effect of Promise Zones on the income distribution.  These estimates are derived from event-study regressions using the baseline specification on the share of households in different income buckets that are shown in Figure \ref{fig:comp_inc} \footnote{Due to the data availability only in nominal terms, Panel (c) also includes a correction for inflation owing to the slightly different baseline distributions. This correction was derived by assuming underlying income was uniform across the income buckets and simulating the difference-in-difference estimate produced by inflation at the observed levels of the Urban CPI. The confidence interval reflect the additional uncertainty from this procedure}. The roughly 2 percentage point decrease in households making less than \$25,000 a year is consistent with the poverty rates estimate in Panel (a) of Figure \ref{fig:main_plots}. On the other side of the distribution, there was a similar increase of almost 2 percentage points of households making over \$100,000 a year in the Designee neighborhoods. 

\begin{table}[h]
    \centering
    \begin{threeparttable}
        
    \begin{tabular}{lccccc} 
        \toprule
        & Under \$25k & \$25-49k & \$50-74k & \$75-99k & Over \$100k \\ 
        \midrule
        ATT Estimate & -0.027(0.008) & -0.005(0.010) & 0.010(0.004) & 0.005(0.005) & 0.019(0.006) \\ 
        Baseline share & 0.444 & 0.289 & 0.139 & 0.063 & 0.065 \\ 
        \bottomrule
    \end{tabular}
    \caption{Change in Income Distribution Relative to Initial Shares}
    \label{tab:comp_inc}
    \begin{tablenotes}
        \small 
        \item Notes: ATT estimates calculated over event times 4-7. \cite{conley_gmm_1999} spatial kernel standard errors with 10 mile cutoff in parentheses.
    \end{tablenotes}
    \end{threeparttable}

\end{table}

Table \ref{tab:comp_inc} shows the inflation-corrected ATT estimates for each group, with the groups over \$100k aggregated together, along with the pre-period share of households in each group in the Designee neighborhoods. Since the program targeted the poorest residents in these areas, it seems unlikely that it would have differentially propelled almost 30\% of the people in the \$75-\$99k group up into the \$100k+ group. For that to be consistent with the other results, it would also mean that an additional 17\% of the \$50k-\$74k group moved up, an additional 12\% of the \$25k-\$49k group moved up, and an additional 7\% of the under \$25k group moved up. A more likely explanation for the patterns seen here are that poorer residents were differentially moving out of these neighborhoods, being replaced by richer residents that were differentially moving into these neighborhoods. There is no differential change in population, so these results are unchanged when plotted in the counts of people in each category (see Appendix Figure \ref{fig:comp_inc_tot}).

\subsubsection{Demographic Characteristics}

\begin{figure}[h]
\centering
    \subfloat[\% Non-white \label{fig:comp_nonwhite}]{\includegraphics[width=0.45\linewidth]{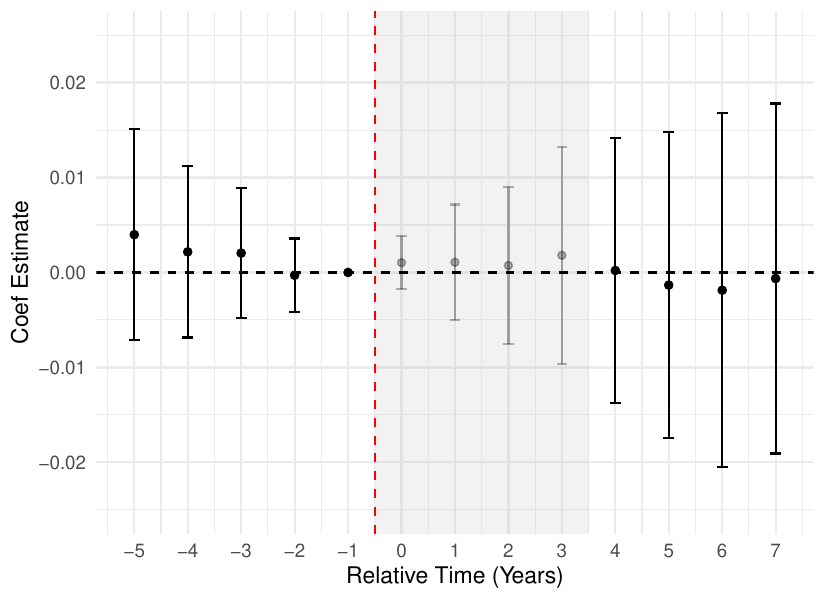}}
    \subfloat[\% Single Parent HH \label{fig:comp_singleparent}]{\includegraphics[width=0.45\linewidth]{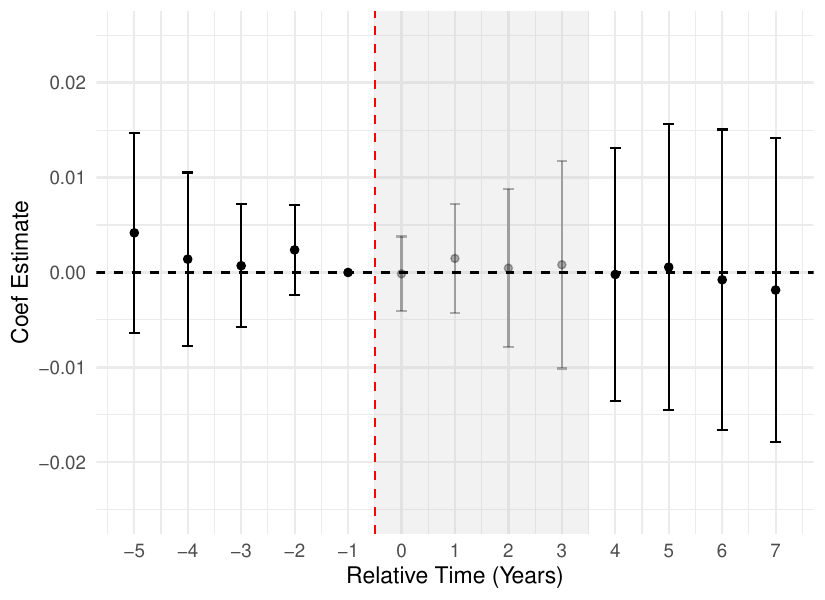}}
    \caption{Effects on Demographic Makeup}
    \label{fig:comp_demo}

    \raggedright
       {\footnotesize Notes: Event study coefficients from baseline empirical specification on different demographic outcomes using \cite{conley_gmm_1999} standard errors. }
\end{figure}

Another approach to investigate changing neighborhood composition is to look into fixed demographic characteristics that cannot change due to this policy. Differences in these outcomes would indicate that the program changed the types of people residing in these neighborhoods. Figure \ref{fig:comp_demo} shows the results from running the baseline specification on two such variables. Panel (a) shows that this policy induced no change in the share of non-white population in these neighborhoods. The confidence intervals of the ATT estimate reject changes larger than 1.7 percentage points in either direction, which are small changes relative to the baseline mean of 85\%. Similarly, it seems extremely unlikely this policy would have cause households that initially were headed by a single parent to transition to dual-headed households. Panel (b) shows that there was no effect of the Promise Zone program on the share of single-parent households in these neighborhoods, and the confidence intervals can reject changes larger than 1.5 percentage points.

\subsection{Change in Federal Funding}

If the main results were not fully driven by a change in neighborhood composition, the most
likely channel for the policy to have impacted the neighborhoods is via an increase in federal
funding. The combination of working with federal liaisons to find grant opportunities, increased manpower from the Americorp VISTA volunteers, and the bonus points on applications for Promise Zone-affiliated federal grant programs should result in designees seeing more federal funding than their finalist counterparts. 


Unfortunately, the geographic target of most federal spending data isn't granular enough to assign it to a Promise Zone or a finalist zone. The consolidated federal spending data offers the addresses of recipients, but the location of a recipient's office often does not correspond to the location where the money is spent. Promise Zones are required to submit Quarterly Reports with the grants that they applied for won, and my access to this data is pending an additional FOIA request. This data should provide a ground truth measure for the designees in the post-period, which can then be used to inform cleaning a federal spending data series to create a credible estimate of the change in federal spending due to this program.

For now, an estimate for the change in federal spending in these areas can be made utilizing the fact that two of the largest Promise Zone-affiliated programs, ones in which they receive preference points for, indicate their grant recipients with precise geographic granularity. The Promise Neighborhood program awards specific schools \$30 million towards ``cradle to career" strategies to improve outcomes for students, and the Choice Neighborhood program awards public housing projects funds to plan and implement the redevelopment of those sites. These two programs list their recipients and award amounts on their own websites, and crucially show the exact schools or public housing projects they target. From these two programs, a back-of-the-envelope population-adjusted difference-in-difference estimate suggests that designees saw a \$148 per-person increase in spending over the 7 year post period. Crude estimates using the addresses of grant recipients from Promise Zone-affiliated programs suggest that these two programs accounted for approximately 5\% of the pre-period spending. This exercise suggests there was a modest increase in federal funding due to the Promise Zone designation.

\section{Conclusion} \label{sec:conclusion}

Despite the large body of evidence demonstrating the effects neighborhoods have on their residents and the places around them, we still lack effective, evidence-backed policies to improve them. This paper makes progress towards that goal by evaluating the federal Promise Zone program, which attempted a ``bottom up" strategy of neighborhood improvement through benefits that increased collaboration between the federal government and local representatives. Using the unsuccessful finalist applicants as a comparison group in an event study framework, this paper finds that the program had significant impacts on neighborhood conditions. Poverty rates decreased, median household incomes increased, and employment to population ratios increased in neighborhoods located in designated Promise Zones. The program's impact was concentrated almost entirely in the last round of designations where the applications were the strongest, and the effects were greater in neighborhoods that were initially lower status. The measured positive impact of the program appears to be driven partly, but not wholly, by a change in neighborhood composition. The effects of the program were not contained to neighborhoods inside its borders; neighborhoods just outside the Promise Zones also saw improvement relative to neighborhoods just outside finalist zones, with effect sizes about half as large as those for neighborhoods inside the borders. This suggests that the program effectively mitigated challenging conditions within Promise Zone neighborhoods that had previously affected surrounding areas.

Comparisons to similar federal place-based policies that designate an area to receive federal benefits suggest that this policy design was more effective at reducing poverty than location-based tax incentives such as in Empowerment Zones, Opportunity Zones, or the New Markets Tax Credit. Given that the other programs entailed significant tax benefits, while Promise Zones only required providing liaisons and VISTA volunteers, the Promise Zone program appears to be a significantly more cost-effective policy to reduce poverty than the aforementioned alternative place-based policies. Taking the stipends of VISTA volunteers (approximately \$25k) and the average pay of a HUD employee (approximately \$110k) puts program costs at about \$3.3 M a year; for comparison, the Empowerment Zone program cost a nominal \$200M in tax credits and \$400M in block grants over the first 6 years of the program \citep{busso_assessing_2013}. Of course, there are indirect costs to the Promise Zone program in terms of the counterfactual winners of grant applications in which the preference points made the difference.

This work provides a starting point for understanding the effectiveness of the Promise Zone program but leaves many avenues open for future research. One of the motivations for the program was to increase economic opportunities for children, and thus the short- and long-run impacts of this program on the children of these neighborhoods are important for understanding the total impact. Additional work with individual-level data would be useful to better understand the degree to which the changing composition accounts for the observed impacts.

\clearpage
\singlespacing
\setlength\bibsep{0pt}
\bibliographystyle{chicago}
\bibliography{references}

\begin{thebibliography}{}

\bibitem[\protect\citeauthoryear{Arefeva, Davis, Ghent, and Park}{Arefeva
  et~al.}{2024}]{arefeva_effect_2024}
Arefeva, A., M.~A. Davis, A.~C. Ghent, and M.~Park (2024, September).
\newblock The {Effect} of {Capital} {Gains} {Taxes} on {Business} {Creation}
  and {Employment}: {The} {Case} of {Opportunity} {Zones}.
\newblock {\em Management Science\/}.
\newblock Publisher: INFORMS.

\bibitem[\protect\citeauthoryear{Arkhangelsky, Athey, Hirshberg, Imbens, and
  Wager}{Arkhangelsky et~al.}{2021}]{arkhangelsky_synthetic_2021}
Arkhangelsky, D., S.~Athey, D.~A. Hirshberg, G.~W. Imbens, and S.~Wager (2021,
  December).
\newblock Synthetic {Difference}-in-{Differences}.
\newblock {\em American Economic Review\/}~{\em 111\/}(12), 4088--4118.

\bibitem[\protect\citeauthoryear{Atkins, Hernández-Lagos, Jara-Figueroa, and
  Seamans}{Atkins et~al.}{2023}]{atkins_jue_2023}
Atkins, R. M.~B., P.~Hernández-Lagos, C.~Jara-Figueroa, and R.~Seamans (2023,
  July).
\newblock {JUE} {Insight}: {What} is the impact of opportunity zones on job
  postings?
\newblock {\em Journal of Urban Economics\/}~{\em 136}, 103545.

\bibitem[\protect\citeauthoryear{Austin, Glaeser, and Summers}{Austin
  et~al.}{2018}]{austin_saving_2018}
Austin, B., E.~Glaeser, and L.~Summers (2018).
\newblock Saving the {Heartland}: {Place}-{Based} {Policies} in 21st {Century}
  {America}.
\newblock {\em Brookings Papers on Economic Activity\/}.

\bibitem[\protect\citeauthoryear{Bartik}{Bartik}{2020}]{bartik_broadening_2020}
Bartik, T. (2020, November).
\newblock Broadening {Place}-{Based} {Jobs} {Policies}: {How} to {Both}
  {Target} {Job} {Creation} and {Broaden} its {Reach}.
\newblock {\em Upjohn Institute Policy Papers\/}.

\bibitem[\protect\citeauthoryear{Borusyak, Jaravel, and Spiess}{Borusyak
  et~al.}{2024}]{borusyak_revisiting_2024}
Borusyak, K., X.~Jaravel, and J.~Spiess (2024, November).
\newblock Revisiting {Event}-{Study} {Designs}: {Robust} and {Efficient}
  {Estimation}.
\newblock {\em The Review of Economic Studies\/}~{\em 91\/}(6), 3253--3285.

\bibitem[\protect\citeauthoryear{Busso, Gregory, and Kline}{Busso
  et~al.}{2013}]{busso_assessing_2013}
Busso, M., J.~Gregory, and P.~Kline (2013, April).
\newblock Assessing the {Incidence} and {Efficiency} of a {Prominent} {Place}
  {Based} {Policy}.
\newblock {\em American Economic Review\/}~{\em 103\/}(2), 897--947.

\bibitem[\protect\citeauthoryear{Callaway and Sant’Anna}{Callaway and
  Sant’Anna}{2021}]{callaway_difference--differences_2021}
Callaway, B. and P.~H.~C. Sant’Anna (2021, December).
\newblock Difference-in-{Differences} with multiple time periods.
\newblock {\em Journal of Econometrics\/}~{\em 225\/}(2), 200--230.

\bibitem[\protect\citeauthoryear{Cengiz, Dube, Lindner, and Zipperer}{Cengiz
  et~al.}{2019}]{cengiz_effect_2019}
Cengiz, D., A.~Dube, A.~Lindner, and B.~Zipperer (2019, August).
\newblock The {Effect} of {Minimum} {Wages} on {Low}-{Wage} {Jobs}*.
\newblock {\em The Quarterly Journal of Economics\/}~{\em 134\/}(3),
  1405--1454.

\bibitem[\protect\citeauthoryear{Chetty and Hendren}{Chetty and
  Hendren}{2018}]{chetty_impacts_2018}
Chetty, R. and N.~Hendren (2018, August).
\newblock The {Impacts} of {Neighborhoods} on {Intergenerational} {Mobility}
  {I}: {Childhood} {Exposure} {Effects}*.
\newblock {\em The Quarterly Journal of Economics\/}~{\em 133\/}(3),
  1107--1162.

\bibitem[\protect\citeauthoryear{Chetty, Hendren, and Katz}{Chetty
  et~al.}{2016}]{chetty_effects_2016}
Chetty, R., N.~Hendren, and L.~F. Katz (2016, April).
\newblock The {Effects} of {Exposure} to {Better} {Neighborhoods} on
  {Children}: {New} {Evidence} from the {Moving} to {Opportunity} {Experiment}.
\newblock {\em American Economic Review\/}~{\em 106\/}(4), 855--902.

\bibitem[\protect\citeauthoryear{Chyn}{Chyn}{2018}]{chyn_moved_2018}
Chyn, E. (2018, October).
\newblock Moved to {Opportunity}: {The} {Long}-{Run} {Effects} of {Public}
  {Housing} {Demolition} on {Children}.
\newblock {\em American Economic Review\/}~{\em 108\/}(10), 3028--3056.

\bibitem[\protect\citeauthoryear{Chyn, Haggag, and Stuart}{Chyn
  et~al.}{2023}]{chyn_effects_2023}
Chyn, E., K.~Haggag, and B.~A. Stuart (2023).
\newblock The {Effects} of {Racial} {Segregation} on {Intergenerational}
  {Mobility}: {Evidence} from {Historical} {Railroad} {Placement}.

\bibitem[\protect\citeauthoryear{Chyn and Katz}{Chyn and
  Katz}{2021}]{chyn_neighborhoods_2021}
Chyn, E. and L.~F. Katz (2021, November).
\newblock Neighborhoods {Matter}: {Assessing} the {Evidence} for {Place}
  {Effects}.
\newblock {\em Journal of Economic Perspectives\/}~{\em 35\/}(4), 197--222.

\bibitem[\protect\citeauthoryear{Conley}{Conley}{1999}]{conley_gmm_1999}
Conley, T.~G. (1999, September).
\newblock {GMM} estimation with cross sectional dependence.
\newblock {\em Journal of Econometrics\/}~{\em 92\/}(1), 1--45.

\bibitem[\protect\citeauthoryear{Diamond and McQuade}{Diamond and
  McQuade}{2019}]{diamond_who_2019}
Diamond, R. and T.~McQuade (2019, June).
\newblock Who {Wants} {Affordable} {Housing} in {Their} {Backyard}? {An}
  {Equilibrium} {Analysis} of {Low}-{Income} {Property} {Development}.
\newblock {\em Journal of Political Economy\/}~{\em 127\/}(3), 1063--1117.
\newblock Publisher: The University of Chicago Press.

\bibitem[\protect\citeauthoryear{Freedman}{Freedman}{2012}]{freedman_teaching_2012}
Freedman, M. (2012, December).
\newblock Teaching new markets old tricks: {The} effects of subsidized
  investment on low-income neighborhoods.
\newblock {\em Journal of Public Economics\/}~{\em 96\/}(11), 1000--1014.

\bibitem[\protect\citeauthoryear{Freedman, Khanna, and Neumark}{Freedman
  et~al.}{2023}]{freedman_jue_2023}
Freedman, M., S.~Khanna, and D.~Neumark (2023, January).
\newblock {JUE} {Insight}: {The} {Impacts} of {Opportunity} {Zones} on {Zone}
  {Residents}.
\newblock {\em Journal of Urban Economics\/}~{\em 133}, 103407.

\bibitem[\protect\citeauthoryear{Haltiwanger, Kutzbach, Palloni, Pollakowski,
  Staiger, and Weinberg}{Haltiwanger et~al.}{2024}]{haltiwanger_children_2024}
Haltiwanger, J.~C., M.~J. Kutzbach, G.~Palloni, H.~O. Pollakowski, M.~Staiger,
  and D.~H. Weinberg (2024, November).
\newblock The children of {HOPE} {VI} demolitions: {National} evidence on labor
  market outcomes.
\newblock {\em Journal of Public Economics\/}~{\em 239}, 105188.

\bibitem[\protect\citeauthoryear{Hanson}{Hanson}{2009}]{hanson_local_2009}
Hanson, A. (2009, November).
\newblock Local employment, poverty, and property value effects of
  geographically-targeted tax incentives: {An} instrumental variables approach.
\newblock {\em Regional Science and Urban Economics\/}~{\em 39\/}(6), 721--731.

\bibitem[\protect\citeauthoryear{Hanson and Rohlin}{Hanson and
  Rohlin}{2013}]{hanson_spatially_2013}
Hanson, A. and S.~Rohlin (2013, January).
\newblock Do spatially targeted redevelopment programs spillover?
\newblock {\em Regional Science and Urban Economics\/}~{\em 43\/}(1), 86--100.

\bibitem[\protect\citeauthoryear{HUD}{HUD}{a}]{hud_promise_nodate}
HUD.
\newblock Promise {Zones} {FY14} {Federal} {Programs}.
\newblock Technical report.

\bibitem[\protect\citeauthoryear{HUD}{HUD}{b}]{hud_promise_nodate-1}
HUD.
\newblock Promise {Zones} {Overview}.

\bibitem[\protect\citeauthoryear{HUD}{HUD}{2014}]{hud_promise_2014}
HUD (2014, January).
\newblock Promise {Zones} 2013 {FAQs}.
\newblock Technical report.

\bibitem[\protect\citeauthoryear{Kitchens and Wallace}{Kitchens and
  Wallace}{2022}]{kitchens_impact_2022}
Kitchens, C. and C.~T. Wallace (2022, July).
\newblock The impact of place-based poverty relief: {Evidence} from the
  {Federal} {Promise} {Zone} {Program}.
\newblock {\em Regional Science and Urban Economics\/}~{\em 95}, 103735.

\bibitem[\protect\citeauthoryear{Kline and Moretti}{Kline and
  Moretti}{2014}]{kline_people_2014}
Kline, P. and E.~Moretti (2014).
\newblock People, {Places}, and {Public} {Policy}: {Some} {Simple} {Welfare}
  {Economics} of {Local} {Economic} {Development} {Programs}.
\newblock {\em Annual Review of Economics\/}~{\em 6\/}(1), 629--662.
\newblock \_eprint: https://doi.org/10.1146/annurev-economics-080213-041024.

\bibitem[\protect\citeauthoryear{Kling, Liebman, and Katz}{Kling
  et~al.}{2007}]{kling_experimental_2007}
Kling, J.~R., J.~B. Liebman, and L.~F. Katz (2007).
\newblock Experimental {Analysis} of {Neighborhood} {Effects}.
\newblock {\em Econometrica\/}~{\em 75\/}(1), 83--119.
\newblock \_eprint:
  https://onlinelibrary.wiley.com/doi/pdf/10.1111/j.1468-0262.2007.00733.x.

\bibitem[\protect\citeauthoryear{Ludwig, Duncan, Gennetian, Katz, Kessler,
  Kling, and Sanbonmatsu}{Ludwig et~al.}{2013}]{ludwig_long-term_2013}
Ludwig, J., G.~J. Duncan, L.~A. Gennetian, L.~F. Katz, R.~C. Kessler, J.~R.
  Kling, and L.~Sanbonmatsu (2013, May).
\newblock Long-{Term} {Neighborhood} {Effects} on {Low}-{Income} {Families}:
  {Evidence} from {Moving} to {Opportunity}.
\newblock {\em American Economic Review\/}~{\em 103\/}(3), 226--231.

\bibitem[\protect\citeauthoryear{Manson, Schroeder, Van~Riper, Knowles, Kugler,
  Roberts, and Ruggles}{Manson et~al.}{2023}]{manson_ipums_2023}
Manson, S., J.~Schroeder, D.~Van~Riper, K.~Knowles, T.~Kugler, F.~Roberts, and
  S.~Ruggles (2023).
\newblock {IPUMS} {National} {Historical} {Geographic} {Information} {System}.

\bibitem[\protect\citeauthoryear{Marsella}{Marsella}{2024}]{marsella_effect_2024}
Marsella, A. (2024, July).
\newblock The effect of the {West} {Philadelphia} {Promise} {Zone} on violent
  crime.

\bibitem[\protect\citeauthoryear{Mayer and Jencks}{Mayer and
  Jencks}{1989}]{mayer_growing_1989}
Mayer, S.~E. and C.~Jencks (1989).
\newblock Growing up in poor neighborhoods: {How} much does it matter?
\newblock {\em Science\/}~{\em 243\/}(4897), 1441--1445.
\newblock Place: US Publisher: American Assn for the Advancement of Science.

\bibitem[\protect\citeauthoryear{Neumark and Young}{Neumark and
  Young}{2019}]{neumark_enterprise_2019}
Neumark, D. and T.~Young (2019, September).
\newblock Enterprise zones, poverty, and labor market outcomes: {Resolving}
  conflicting evidence.
\newblock {\em Regional Science and Urban Economics\/}~{\em 78}, 103462.

\bibitem[\protect\citeauthoryear{Rambachan and Roth}{Rambachan and
  Roth}{2023}]{rambachan_more_2023}
Rambachan, A. and J.~Roth (2023, October).
\newblock A {More} {Credible} {Approach} to {Parallel} {Trends}.
\newblock {\em The Review of Economic Studies\/}~{\em 90\/}(5), 2555--2591.

\bibitem[\protect\citeauthoryear{Reynolds and Rohlin}{Reynolds and
  Rohlin}{2015}]{reynolds_effects_2015}
Reynolds, C.~L. and S.~M. Rohlin (2015, July).
\newblock The effects of location-based tax policies on the distribution of
  household income: {Evidence} from the federal {Empowerment} {Zone} program.
\newblock {\em Journal of Urban Economics\/}~{\em 88}, 1--15.

\bibitem[\protect\citeauthoryear{Roback}{Roback}{1982}]{roback_wages_1982}
Roback, J. (1982).
\newblock Wages, {Rents}, and the {Quality} of {Life}.
\newblock {\em Journal of Political Economy\/}~{\em 90\/}(6), 1257--1278.
\newblock Publisher: University of Chicago Press.

\bibitem[\protect\citeauthoryear{Rosen}{Rosen}{1979}]{rosen_housing_1979}
Rosen, H.~S. (1979, February).
\newblock Housing decisions and the {U}.{S}. income tax: {An} econometric
  analysis.
\newblock {\em Journal of Public Economics\/}~{\em 11\/}(1), 1--23.

\bibitem[\protect\citeauthoryear{Roth}{Roth}{2022}]{roth_pretest_2022}
Roth, J. (2022, September).
\newblock Pretest with {Caution}: {Event}-{Study} {Estimates} after {Testing}
  for {Parallel} {Trends}.
\newblock {\em American Economic Review: Insights\/}~{\em 4\/}(3), 305--322.

\bibitem[\protect\citeauthoryear{Roth}{Roth}{2024}]{roth_interpreting_2024}
Roth, J. (2024, January).
\newblock Interpreting {Event}-{Studies} from {Recent}
  {Diﬀerence}-in-{Diﬀerences} {Methods}.

\bibitem[\protect\citeauthoryear{Sampson, Morenoff, and Gannon-Rowley}{Sampson
  et~al.}{2002}]{sampson_assessing_2002}
Sampson, R.~J., J.~D. Morenoff, and T.~Gannon-Rowley (2002).
\newblock Assessing "{Neighborhood} {Effects}": {Social} {Processes} and {New}
  {Directions} in {Research}.
\newblock {\em Annual Review of Sociology\/}~{\em 28}, 443--478.
\newblock Publisher: Annual Reviews.

\bibitem[\protect\citeauthoryear{Staiger, Palloni, and Voorheis}{Staiger
  et~al.}{2024}]{staiger_neighborhood_2024}
Staiger, M., G.~Palloni, and J.~Voorheis (2024, October).
\newblock Neighborhood {Revitalization} and {Residential} {Sorting}.

\bibitem[\protect\citeauthoryear{Sun and Abraham}{Sun and
  Abraham}{2021}]{sun_estimating_2021}
Sun, L. and S.~Abraham (2021, December).
\newblock Estimating dynamic treatment effects in event studies with
  heterogeneous treatment effects.
\newblock {\em Journal of Econometrics\/}~{\em 225\/}(2), 175--199.

\bibitem[\protect\citeauthoryear{Tach and Emory}{Tach and
  Emory}{2017}]{tach_public_2017}
Tach, L. and A.~D. Emory (2017, November).
\newblock Public {Housing} {Redevelopment}, {Neighborhood} {Change}, and the
  {Restructuring} of {Urban} {Inequality}.
\newblock {\em American Journal of Sociology\/}~{\em 123\/}(3), 686--739.
\newblock Publisher: The University of Chicago Press.

\bibitem[\protect\citeauthoryear{Wilson}{Wilson}{1987}]{wilson_truly_1987}
Wilson, W.~J. (1987).
\newblock {\em The {Truly} {Disadvantaged}: {The} {Inner} {City}, the
  {Underclass}, and {Public} {Policy}}.
\newblock Chicago: University of Chicago Press.
\newblock Backup Publisher: University of Chicago Press.

\bibitem[\protect\citeauthoryear{Wing, Freedman, and Hollingsworth}{Wing
  et~al.}{2024}]{wing_stacked_2024}
Wing, C., S.~M. Freedman, and A.~Hollingsworth (2024, January).
\newblock Stacked {Difference}-in-{Differences}.

\bibitem[\protect\citeauthoryear{Zapolsky, Ault, Brock, Rekhi, and
  Sweeny}{Zapolsky et~al.}{2019}]{zapolsky_promise_2019}
Zapolsky, S., M.~Ault, J.~Brock, J.~Rekhi, and D.~Sweeny (2019, February).
\newblock Promise {Zones}: {Initial} {Implementation} {Assessment} {Report}.
\newblock Technical report.

\end{thebibliography}

\clearpage

\onehalfspacing

\appendix
\counterwithin{figure}{section}
\counterwithin{table}{section}

\section{Supplementary Tables \& Figures} 
\label{app:tabs}

\begin{table}[H]
\begin{threeparttable}
    
    \centering
    \begin{tabular}{lcccc}
        \hline
        & NSI & Poverty Rate & Med HH Income & Emp-Pop Ratio \\
        \hline
        ATT Estimate & 0.211(0.071) & -0.024(0.010) & 0.090(0.029)& 0.014(0.006)\\
        p-value & 0.00 & 0.02 & 0.00 & 0.01\\
        \hline
    \end{tabular}
    \caption{ATT Estimates in Sun-Abraham Estimator}
    \label{tab:sun_abe}
    \begin{tablenotes}
    \small
    \item Notes: ATT estimates calculated over event times 4-7. \cite{conley_gmm_1999} spatial kernel standard errors with 10 mile cutoff in parentheses. 
    \end{tablenotes}
    \end{threeparttable}

\end{table}

\begin{figure}[H]
    \centering
    \includegraphics[width=0.7\linewidth]{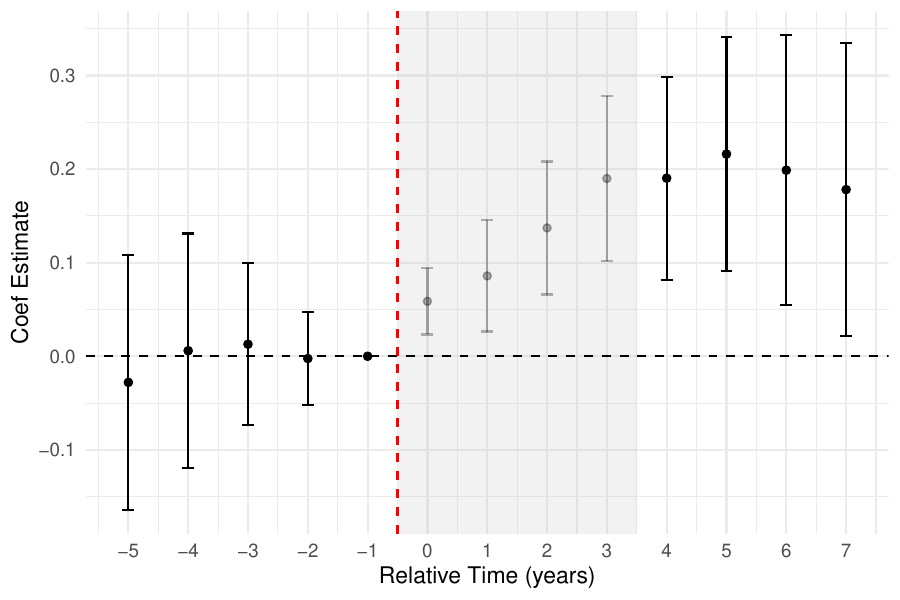}
    \caption{Baseline Specification on Neighborhood Status Index (Scaled Partial Treatment)}
    \label{fig:main_nsi_scaled}
    \raggedright
       {\footnotesize Notes: Event study coefficients from baseline empirical specification on Neighborhood Status Index using \cite{conley_gmm_1999} standard errors. Grayed out areas correspond to partial treatment estimates that have been scaled up to represent full treatment.}
\end{figure}

\begin{figure}[H]
    \centering
    \subfloat[Low-Status Surrounding Areas \label{fig:het_surround_low}]{\includegraphics[width=0.45\linewidth]{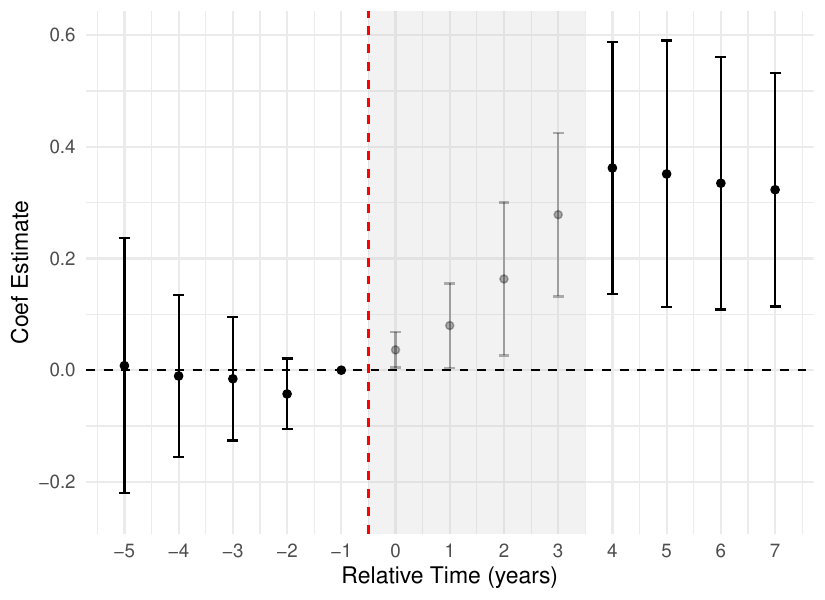}}
    \subfloat[High-Status Surrounding Areas \label{fig:het_surround_high}]{\includegraphics[width=0.45\linewidth]{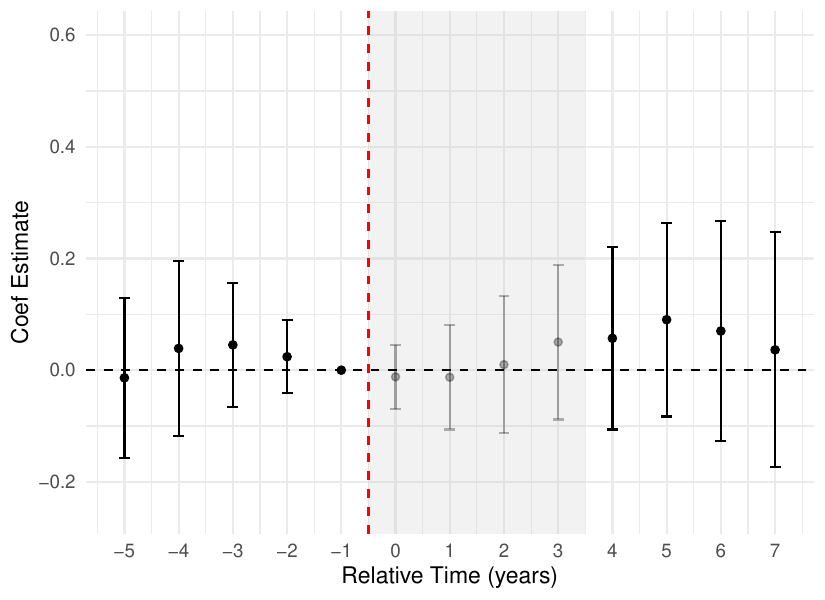}}
    \caption{Treatment Effects by Status of Surrounding Neighborhoods}
    \label{fig:het_surround}
    \raggedright
       {\footnotesize Notes: Event study coefficients from baseline empirical specification on Neighborhood Status Index using \cite{conley_gmm_1999} standard errors. }
\end{figure}

\begin{figure}[H]
    \centering
    \subfloat[Non-Opportunity Zones \label{fig:het_oz_0}]{\includegraphics[width=0.45\linewidth]{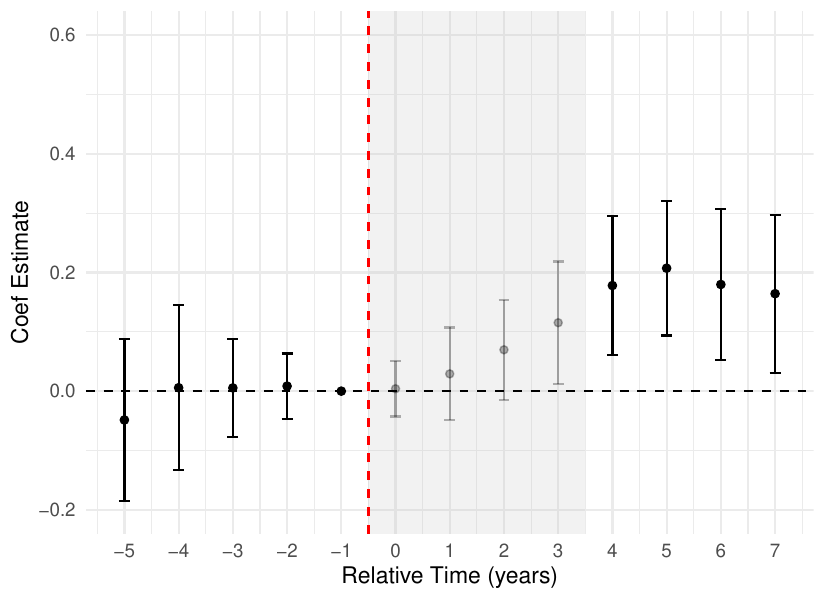}}
    \subfloat[Opportunity Zones \label{fig:het_oz_1}]{\includegraphics[width=0.45\linewidth]{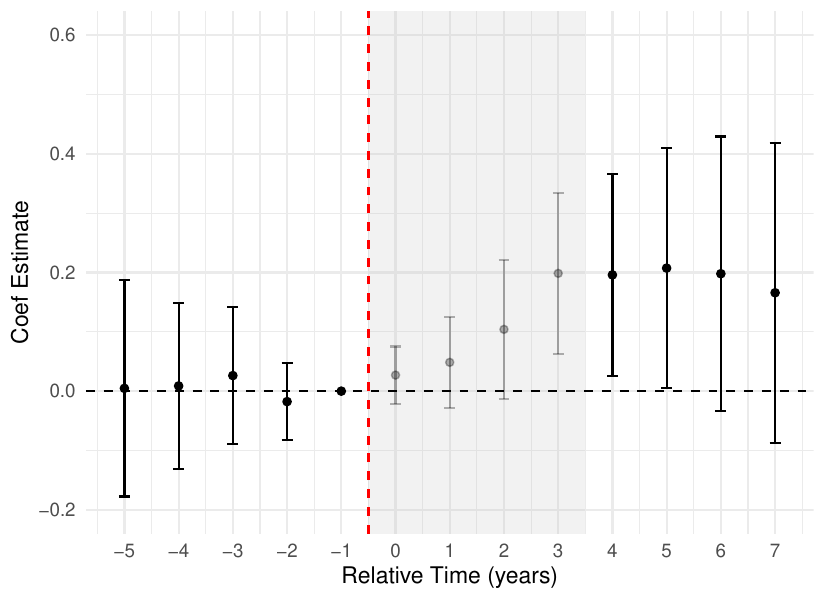}}
    \caption{Treatment Effects by Opportunity Zone Classification}
    \label{fig:het_oz}
    \raggedright
       {\footnotesize Notes: Event study coefficients from baseline empirical specification on Neighborhood Status Index using \cite{conley_gmm_1999} standard errors. Both treatment and control groups restricted to be either Opportunity Zones or not.}
\end{figure}

\begin{figure}[H]
    \centering
    \includegraphics[width=0.6\linewidth]{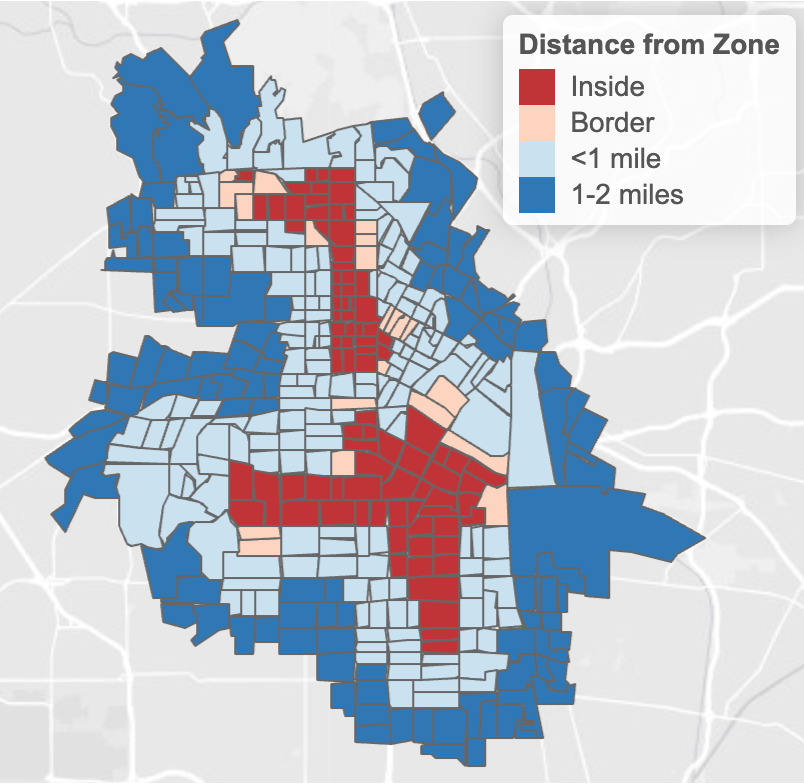}
    \caption{Map of Tracts in each Spillover Group}
    \label{fig:spill_map}

    \raggedright
       {\footnotesize Notes: Map of census tracts in downtown Los Angeles. Tracts in dark red are 75\% or more inside either the LA Promise Zone or South LA Promise Zone. Tracts in light red are between 25\% and 75\% inside a zone. Tracts in light blue are less than 25\% inside a zone and have centroids within 1 mile of a zone border. Tracts in dark blue have centroids between 1-2 miles from zone border.}
\end{figure}

\begin{figure}[H]
    \centering
    \includegraphics[width=0.8\linewidth]{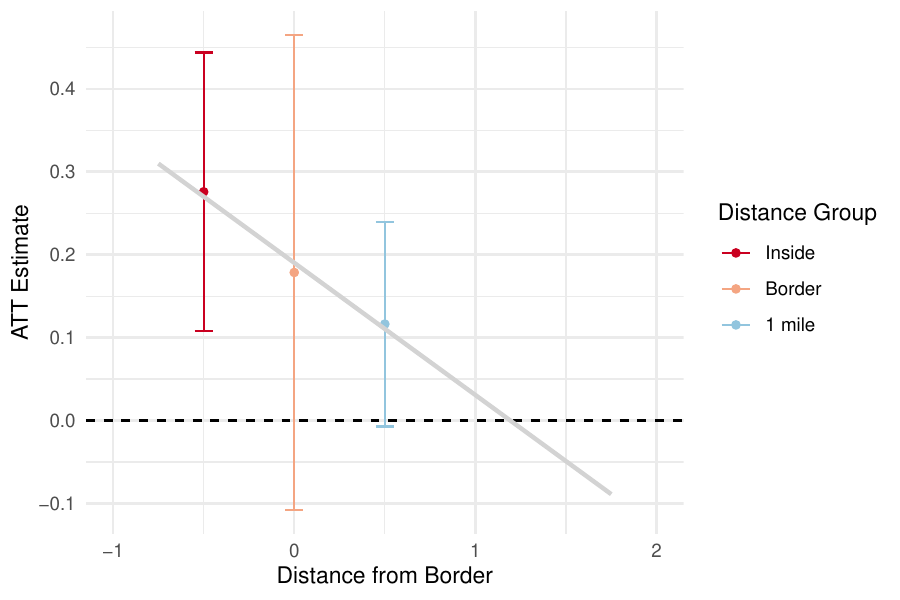}
    \caption{Spatial Decay of Treatment Effects from Border}
    \label{fig:spill_decay}
    \raggedright
       {\footnotesize Notes: ATT estimates from baseline empirical specification run in different distance groups on Neighborhood Status Index using \cite{conley_gmm_1999} standard errors. Groups are plotted at the average tract centroid distance from the zone border. Gray line is linear line of best fit from point estimates. }
\end{figure}

\begin{figure}[h]
    \centering
    \begin{tabular}{ccc}
    \subfloat[Households Share: Under \$25k \label{fig:comp_inc_u25}]{\includegraphics[width=0.3\linewidth]{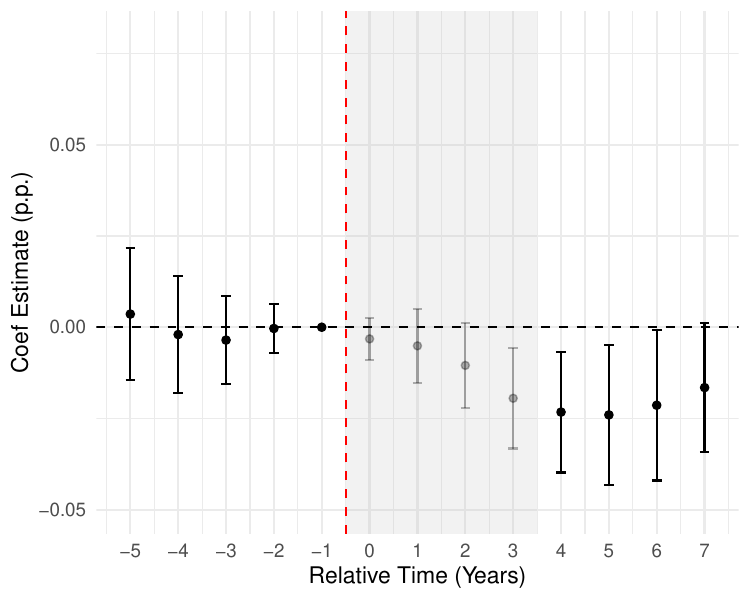}} &
    \subfloat[Households Share: \$25k-\$49k \label{fig:comp_inc_25_49}]{\includegraphics[width=0.3\linewidth]{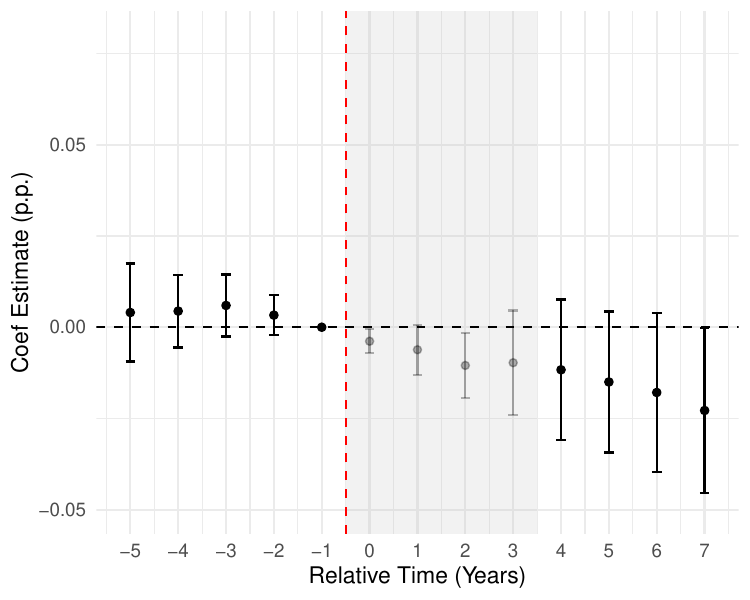}} &
    \subfloat[Households Share: \$50k-\$74k \label{fig:comp_inc_50_74}]{\includegraphics[width=0.3\linewidth]{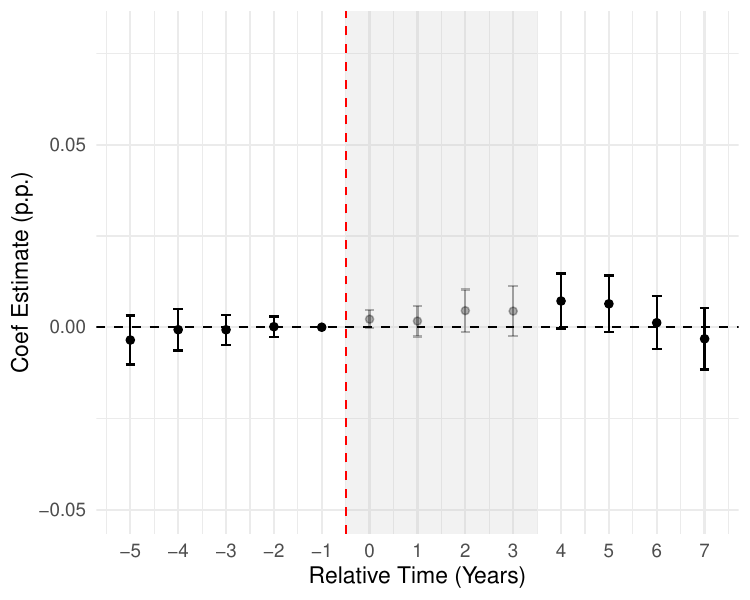}} \\ 
    \subfloat[Households Share: \$75k-\$99k \label{fig:comp_inc_75_99}]{\includegraphics[width=0.3\linewidth]{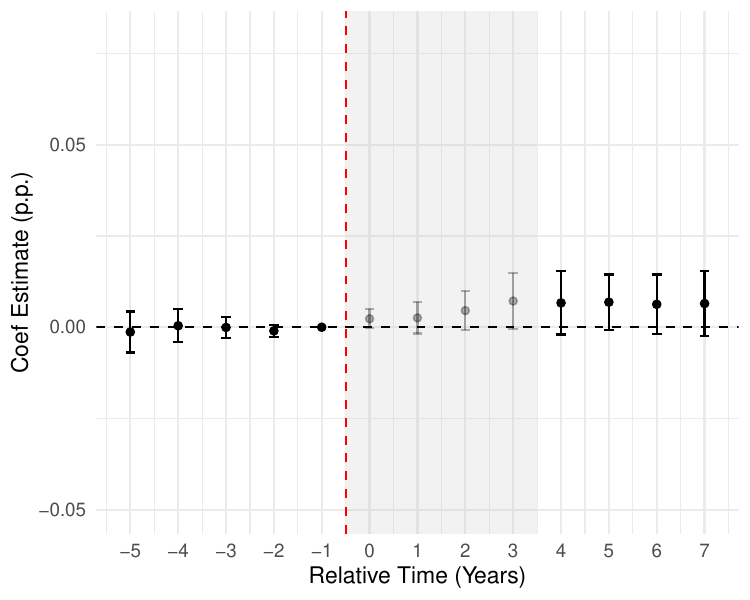}} & 
    \subfloat[Households Share: Over \$100k \label{fig:comp_inc_100_plus}]{\includegraphics[width=0.3\linewidth]{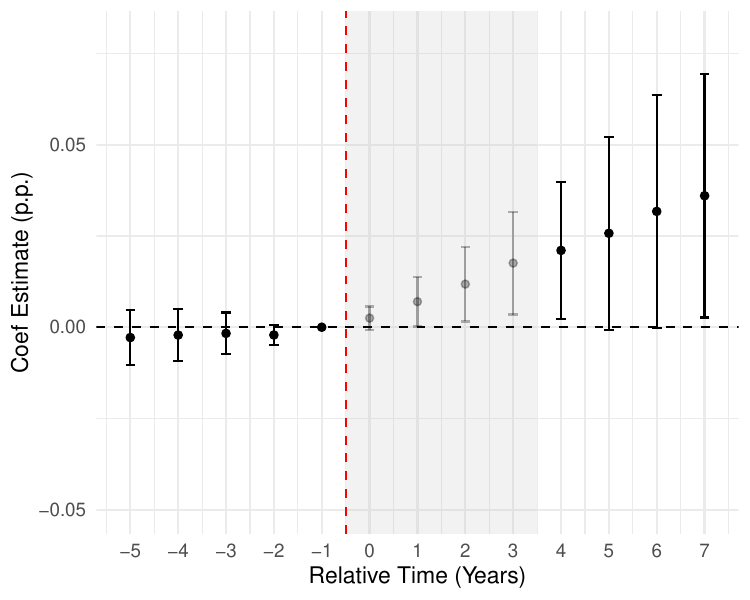}} \\
    \end{tabular}
    \caption{Effects on Relative Shares of Income Distribution}
    \label{fig:comp_inc}
\raggedright
       {\footnotesize Notes: Event study coefficients from baseline empirical specification on share of households in different income buckets using \cite{conley_gmm_1999} standard errors. }
    
\end{figure}

\begin{figure}[H]
    \centering
    \begin{tabular}{ccc}
    \subfloat[Households: Under \$25k \label{fig:comp_inc_u25_tot}]{\includegraphics[width=0.3\linewidth]{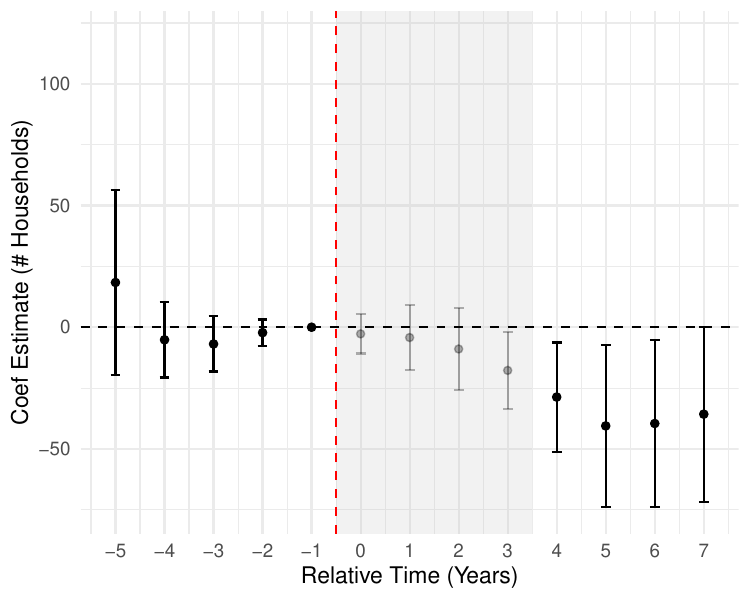}} &
    \subfloat[Households: \$25k-\$49k \label{fig:comp_inc_25_49_tot}]{\includegraphics[width=0.3\linewidth]{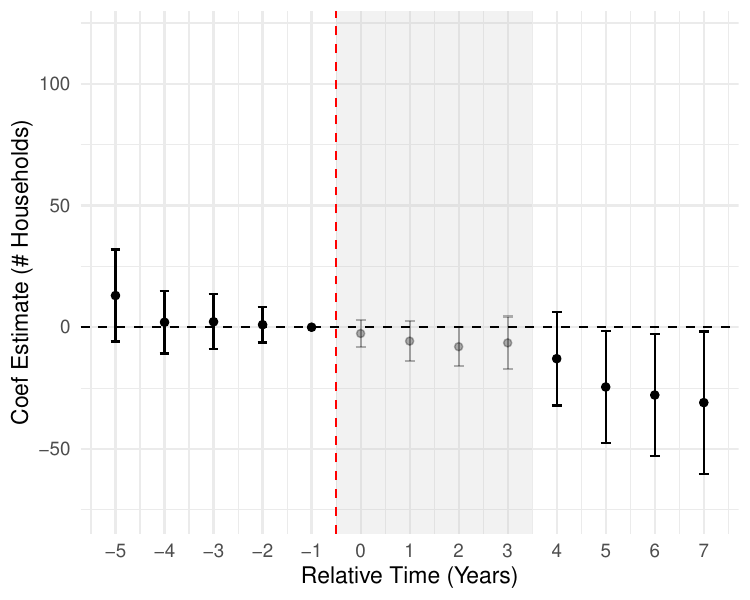}} &
    \subfloat[Households: \$50k-\$74k \label{fig:comp_inc_50_74_tot}]{\includegraphics[width=0.3\linewidth]{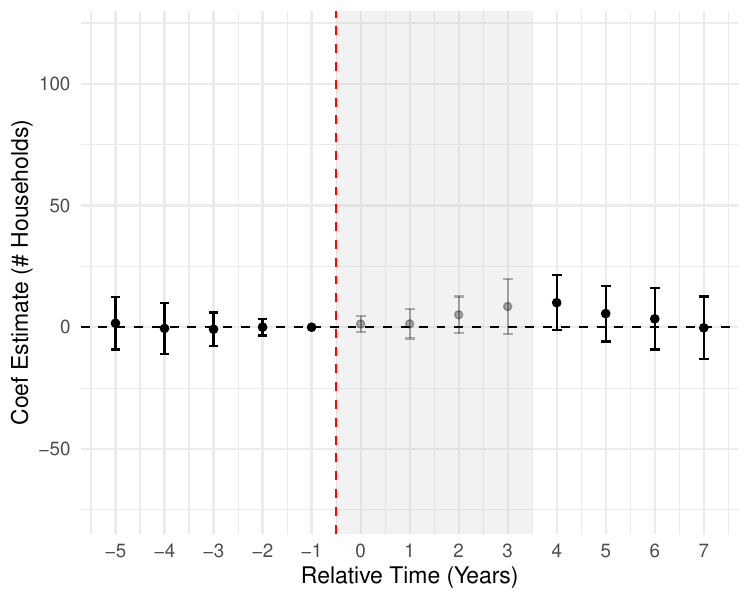}} \\ 
    \subfloat[Households: \$75k-\$99k \label{fig:comp_inc_75_99_tot}]{\includegraphics[width=0.3\linewidth]{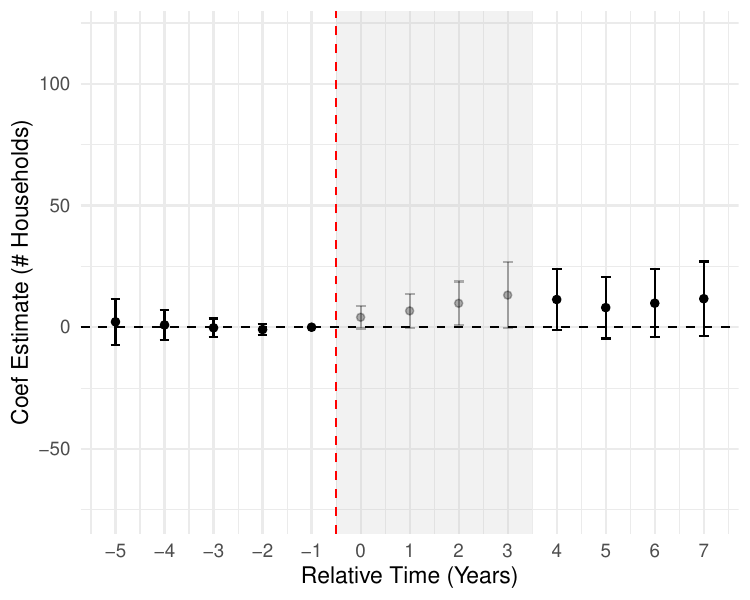}} &
    \subfloat[Households: Over \$100k \label{fig:comp_inc_100_plus_tot}]{\includegraphics[width=0.3\linewidth]{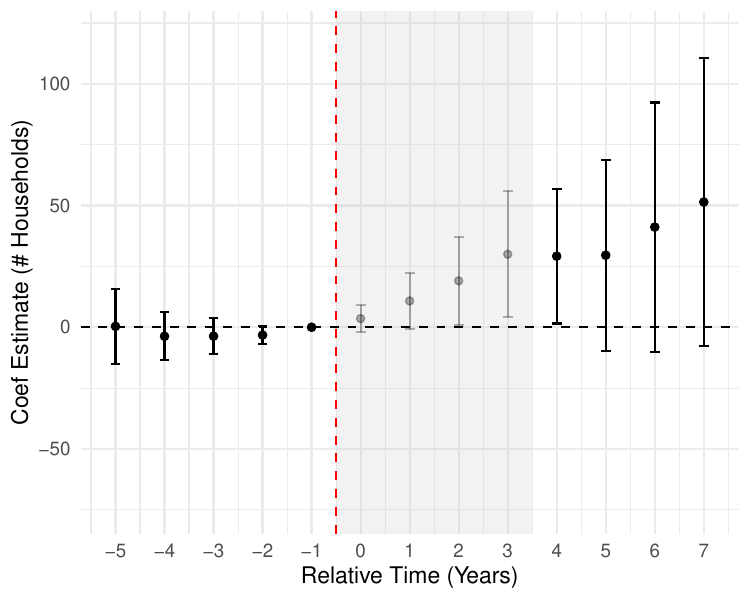}} \\
    \end{tabular}
    \caption{Effects on Counts of Income Distribution}
    \label{fig:comp_inc_tot}

    \raggedright
       {\footnotesize Notes: Event study coefficients from baseline empirical specification on counts of households in different income buckets using \cite{conley_gmm_1999} standard errors.}
\end{figure}

\begin{figure}[H]
    \centering
    \begin{tabular}{cc}
    \subfloat[\% College \label{fig:comp_pct_college}]{\includegraphics[width=0.45\linewidth]{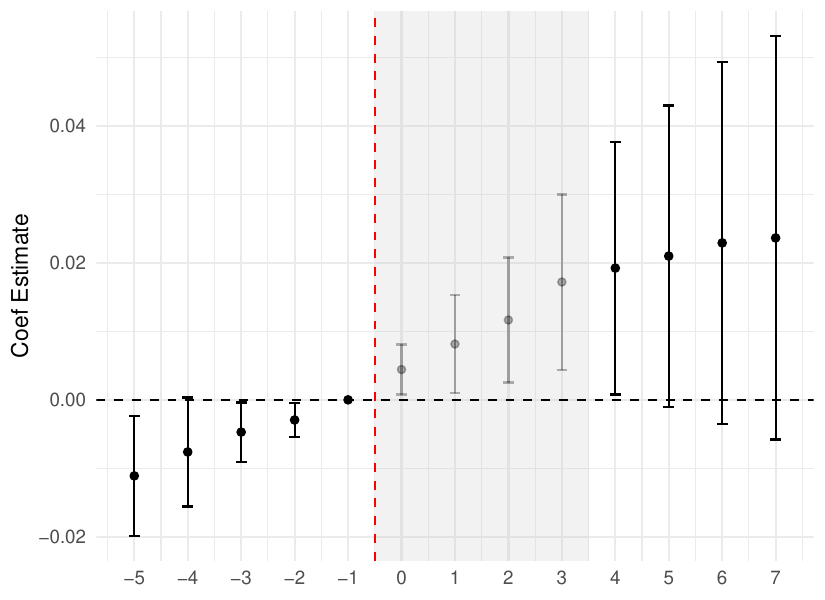}} &
    \subfloat[Median HH Duration in Residence \label{fig:comp_med_dur}]{\includegraphics[width=0.45\linewidth]{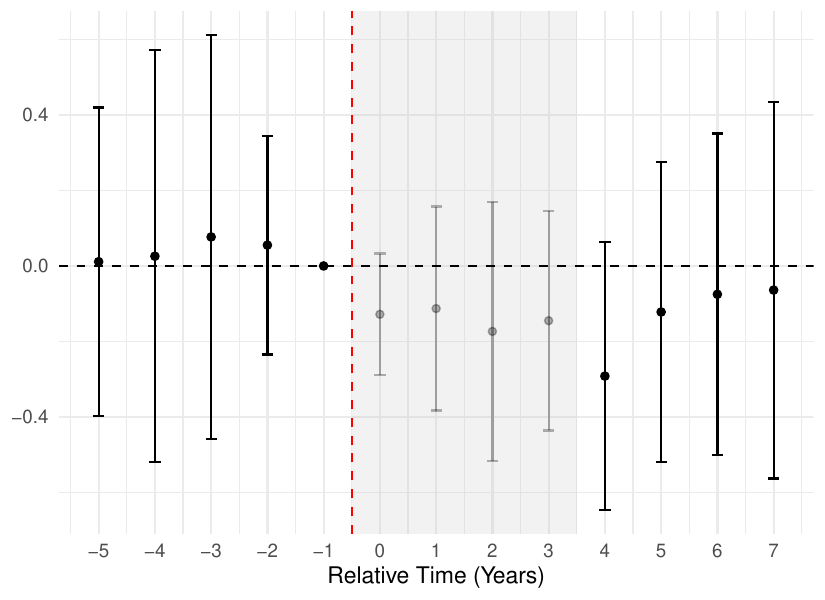}} \\
    \subfloat[Median House Value (log) \label{fig:comp_house_val}]{\includegraphics[width=0.45\linewidth]{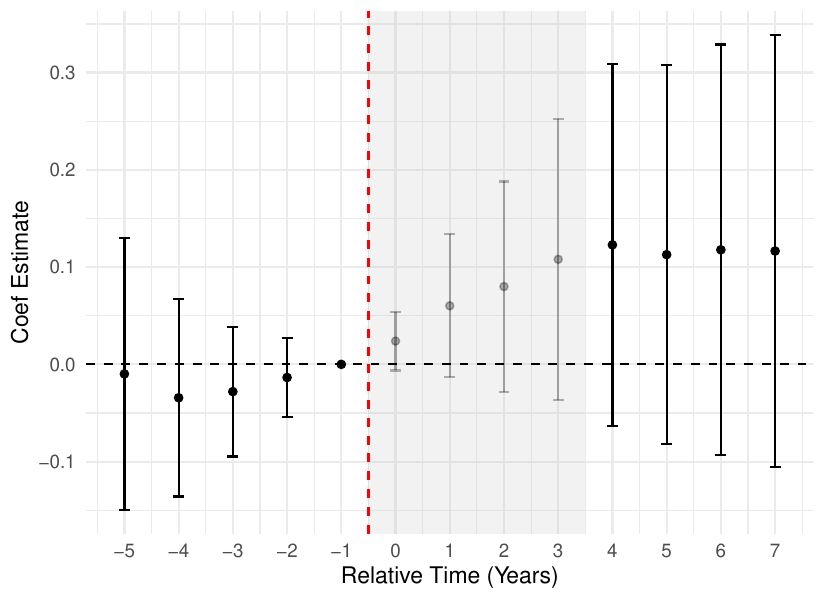}} & 
    \subfloat[Median Rent (log) \label{fig:comp_rent}]{\includegraphics[width=0.45\linewidth]{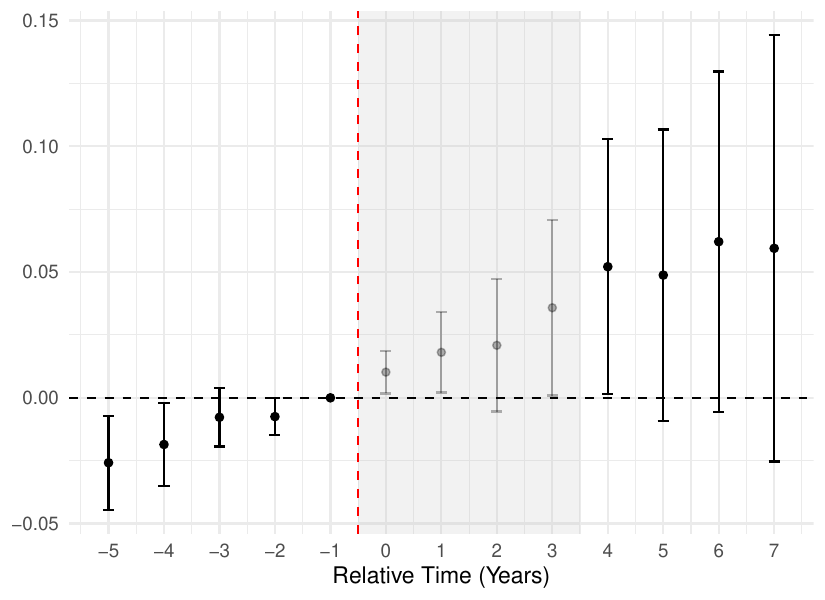}} \\
    \end{tabular}
    \caption{Additional Composition Outcomes}
    \label{fig:comp_ext}
\raggedright
       {\footnotesize Notes: Event study coefficients from baseline empirical specification on various compositional outcomes using \cite{conley_gmm_1999} standard errors. }
    
\end{figure}

\section{Data Construction}
\label{app:data}

Maps for the designees and finalists were obtained as documents showing the proposed borders overlaid on a street map. These images were georeferenced in QGIS, and polygons were drawn on to match the borders. These polygons were then spatially merged with block-level data from the 2010 Census containing population estimates. The share of census blocks that overlapped in area with the zone polygons was multiplied by the block population to get an estimate of the population from that block living within the zone borders. The block-level crosswalk was then aggregated up to the census tract level to get an estimate of the population in each census tract that resided within the zone borders. These ``population in the zone" weights were then used in the baseline regressions to weight the tract observations. 

Census tracts were defined using the 2010 borders. ACS data for years prior to 2010 and after 2020 are provided for census tracts covering different geographic areas. To ensure geographic continuity across years, those estimates were crosswalked to 2010 census tracts using crosswalks provided by the Longitudinal Tract Database. The crosswalking process used only averages of underlying data that were non-missing to ensure as much data coverage as possible.

To ensure a consistent sample, only tracts that formed a balanced panel were kept. Any that had missing data for any of the three primary outcomes in any years in the study period were dropped entirely. 

\section{Additional Analyses}
\subsection{Local Comparison Tracts} \label{sec:local_comps}

This section estimates the effect of the Promise Zone program using nearby neighborhoods as a control group and a synthetic difference-in-difference estimator to reweight them \citep{arkhangelsky_synthetic_2021}. 

For each designated Promise Zone, all tracts with centroids between 2 and 10 miles from the border are taken as donor units. A synthetic difference-in-difference estimator with only unit, but not time, weights is applied to reweight the donor units, and then event study coefficients are generated using a standard two-way dynamic event study specification. The procedure only optimizes over 3 pre-treatment periods, leaving the earlier ones as ``tests" for goodness of fit. For inference, placebo treatment areas are created from the control pool by randomly selecting tracts (as many as were in the treatment group) and running the same synthetic difference-in-difference estimator. This procedure was repeated 100 times, and the standard deviation across placebo estimates was recorded as the standard error for the true estimate.

\begin{figure}[H]
    \centering
    \includegraphics[width=0.8\linewidth]{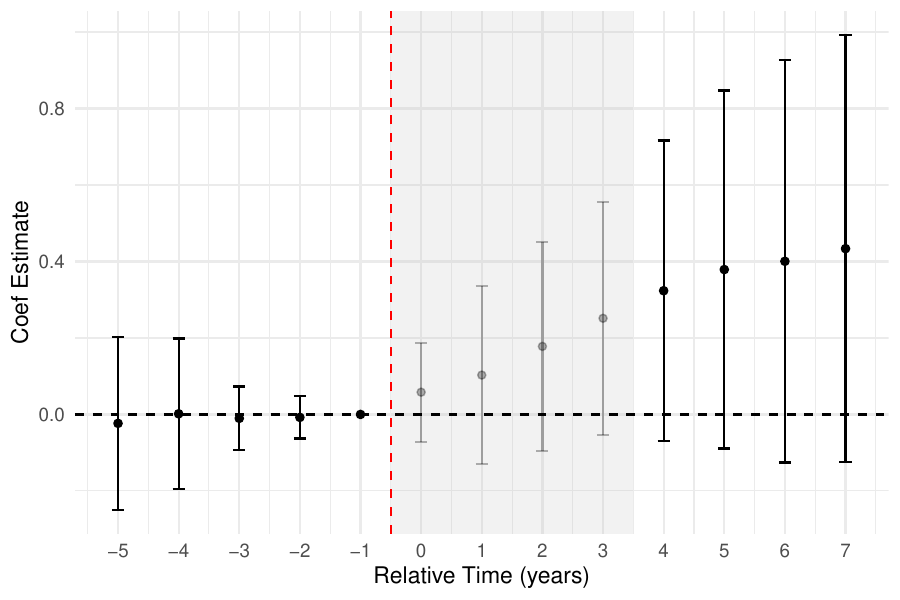}
    \caption{Synthetic DiD Estimates on Neighborhood Status Index}
    \label{fig:sdid_index}
       \raggedright
       {\footnotesize Notes: Event study coefficients from Synthetic Difference-in-Difference procedure described in Section \ref{sec:local_comps} on Neighborhood Status Index. Confidence interval generated from placebo inference procedure. }
\end{figure}

\begin{figure}[H]
    \centering
    \subfloat[Poverty Rate \label{fig:sdid_pov}]{\includegraphics[width=0.33\linewidth]{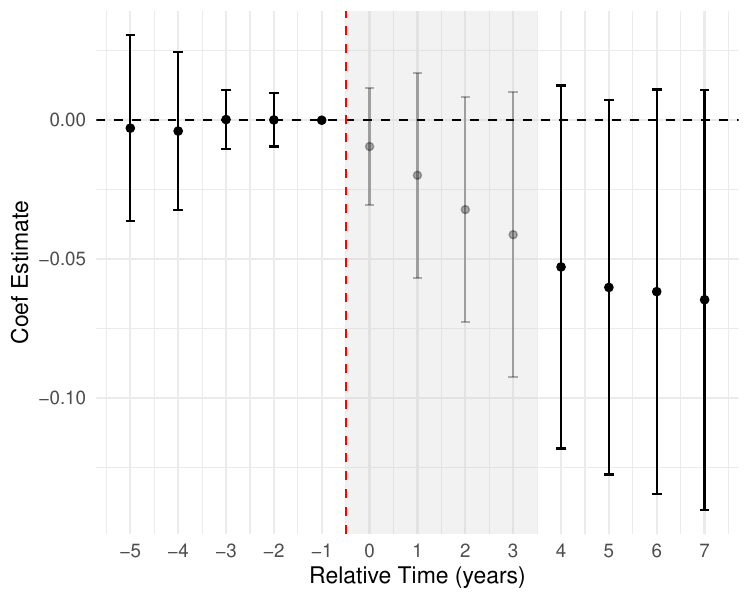}}
    \subfloat[Med HH Income (log) \label{fig:sdid_inc}]{\includegraphics[width=0.33\linewidth]{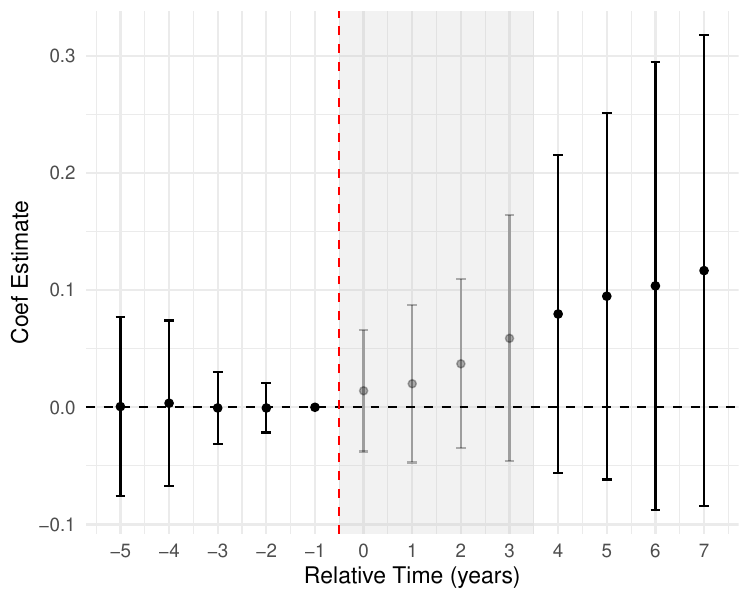}}
    \subfloat[Emp-Pop Ratio \label{fig:sdid_emp}]{\includegraphics[width=0.33\linewidth]{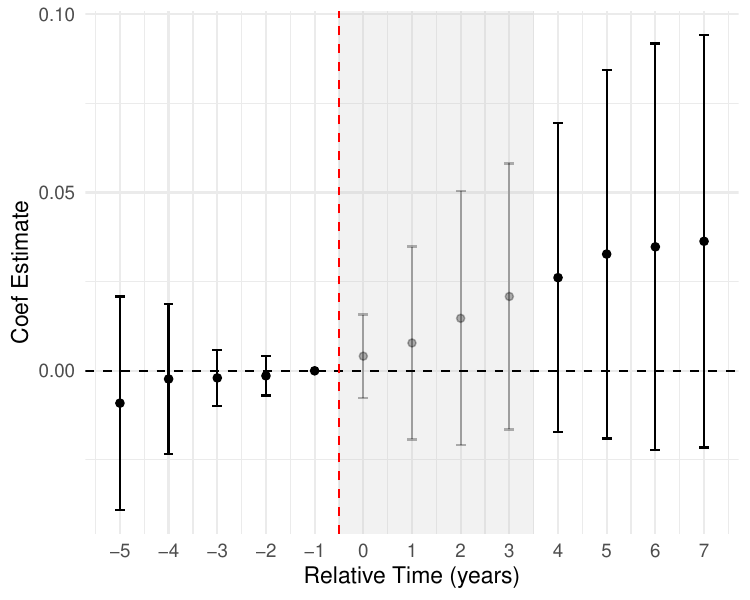}}
    \caption{Synthetic DiD Estimates on Primary Outcomes}
    \label{fig:sdid_main}
        \raggedright
       {\footnotesize Notes: Event study coefficients from Synthetic Difference-in-Difference procedure described in Section \ref{sec:local_comps} on different outcomes. Confidence interval generated from placebo inference procedure. }
\end{figure}

Estimates across the 14 Promise Zones are then aggregated by population weights to create a single event study for each outcome variable. The event study estimates for the Neighborhood Status Index are plotted in Figure \ref{fig:sdid_index}, and the three components of the index in Figure \ref{fig:sdid_main}. This procedure was run separately for all 4 outcomes. These figures show qualitatively similar results to the figures from the main specification in Section \ref{ssec:res_main}. The effect sizes are larger but this procedure results in more imprecise measurements. 

\subsection{Treatment Dynamics} \label{sec:dynamics}

A limitation of the outcome data coming in 5-year estimates is that it's harder to understand the dynamic effects of the treatment over time. This section makes two additional assumptions: that each 5-year estimate is an equal-weighted average of the underlying single-year values, and that parallel trends holds exactly in the pre-period. This second assumption ensures that differences taken with respect to the base year of $t=-1$ are exchangeable with differences taken with respect to other years in the pre-period. With these additional assumptions, it's possible to take differences in coefficients between 5-year estimates to back out single-year changes. For example, the new estimate for $t=1$ is formed by taking the difference between the old estimates in $t=1$ and $t=0$ and multiplying by 5. This difference is now capturing the effects of $t=1$ to the baseline and $t=-4$ relative to the baseline. Under the second assumption of perfect pre-period parallel trends, the second term drops out, so that coefficient is now just the ATT in $t=1$ relative to the baseline of $t=-1$. 

\begin{figure}[h!]
    \centering
    \includegraphics[width=\linewidth]{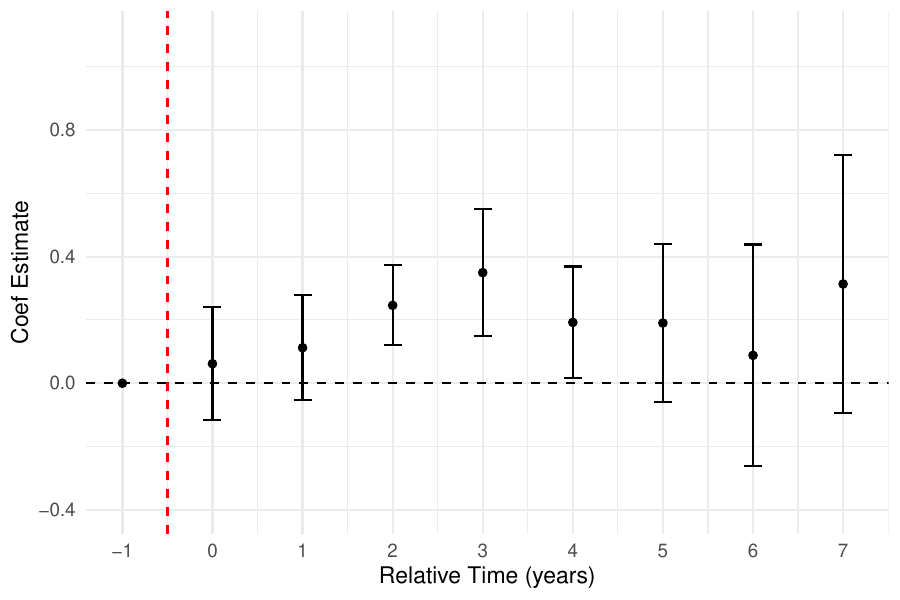}
    \caption{Dynamic Effects of Policy over Time}
    \label{fig:dynamics}
    \raggedright
       {\footnotesize Notes: Event study coefficients constructed as linear combinations of those from baseline empirical specification on Neighborhood Status Index using \cite{conley_gmm_1999} standard errors. }
\end{figure}

The results of this exercise are plotted in Figure \ref{fig:dynamics}. It appears that the policy took a few years to make a difference, but the treatment effect was diminished in later years. It should be noted that the further from the baseline, the more a deviation from perfect parallel trends in the pre-period seems likely. The fact that this exercise shows that treatment effects took a few years to kick in aligns with expectations over a program like this.  The ATT estimate from this procedure is 0.22(se = 0.10), which is also reassuring that it lines up with the main results.

\end{document}